\documentclass[10pt,journal]{IEEEtran}
\IEEEoverridecommandlockouts
\ifCLASSINFOpdf
\else
\fi
\usepackage{bbm}
\usepackage{verbatim}
\usepackage{graphicx}
\usepackage{cite}
\usepackage{url}
\usepackage[cmex10]{amsmath}
\usepackage{amssymb}
\usepackage{amsmath}
\usepackage{dsfont}
\usepackage{algorithm, algorithmicx, algpseudocode}
\usepackage[caption=true,font=footnotesize]{subfig}
\usepackage{color}
\usepackage{cleveref}
\usepackage{cite}
\usepackage{epstopdf}
\usepackage{calc}
\usepackage{array}
\usepackage{lipsum}
\usepackage{siunitx}
\usepackage[printonlyused]{acronym}
\usepackage{amsmath}
\usepackage{multirow}

\DeclareSIUnit{\belmilliwatt}{Bm}
\DeclareSIUnit{\bps}{bps}
\DeclareSIUnit{\dBm}{\deci\belmilliwatt}

\newtheorem{proposition}{Proposition}

\newcommand{\ggroup}[0]{g_{\mathrm{group}}}
\newcommand{\BW}[0]{\mathrm{BW}}
\newcommand{\Tdl}[0]{T_{\mathrm{DL}}}
\newcommand{\Tpacketloss}[0]{T_{\mathrm{packetloss}}}
\newcommand{\Tl}[0]{T_{\mathrm{tot}}}
\newcommand{\Tactivation}[0]{T_{\mathrm{activation}}}
\newcommand{\Tfailure}[0]{T_{\mathrm{failure}}}
\newcommand{\Tsync}[0]{T_{\mathrm{sync}}}
\newcommand{\Tcont}[0]{T_{\mathrm{cont}}}

\newcommand{\bstaru}[0]{b^{\star}_u}
\newcommand{\bs}[0]{\text{B}}
\newcommand{\Pplr}[0]{P_{\mathrm{PLR}}}

\newcommand{\ue}[0]{\text{U}}

\newcommand{\hermitian}[0]{\text{H}}
\newcommand{\transpose}[0]{\text{T}}

\newcommand{\ie}[0]{i.e.}

\acrodef{NB-IoT}{Narrowband Internet of Things}
\acrodef{FFT}{fast Fourier transform}
\acrodef{PN}{pseudorandom noise}
\acrodef{PLR}{packet loss rate}
\acrodef{RF}{radio frequency}
\acrodef{IoT}{internet of things}
\acrodef{eMBB}{enhanced mobile broadband}
\acrodef{IWSN}{industrial/intelligent wireless sensor networks}
\acrodef{MPC}{multipath components}
\acrodef{mmW}{millimeter-wave}
\acrodef{BS}{base station}
\acrodef{RB}{resource block}
\acrodef{UE}{user equipment}
\acrodef{CP}{cyclic prefix}
\acrodef{SOTA}{state of the art}
\acrodef{AoA}{angle of arrival}
\acrodef{AoD}{angle of departure}
\acrodef{AWV}{antenna weight vector}
\acrodef{ADC}{analog-to-digital converter}
\acrodef{BB}{baseband}
\acrodef{LoS}{line-of-sight}
\acrodef{NLoS}{non-line-of-sight}
\acrodef{DFT}{discrete Fourier transform}
\acrodef{SNR}{signal-to-noise ratio}
\acrodef{SINR}{signal-to-interference-plus-noise ratio}
\acrodef{Tx}{transmitter}
\acrodef{AWGN}{additive white Gaussian noise}
\acrodef{Rx}{receiver}
\acrodef{MIMO}{multiple-input multiple-output}
\acrodef{DSP}{digital signal processing}
\acrodef{OFDM}{orthogonal frequency-division multiplexing}
\acrodef{OFDMA}{orthogonal frequency-division multiple access}
\acrodef{TTI}{transmission time interval}
\acrodef{HARQ}{hybrid automatic repeat request}
\acrodef{GF}{grant free}
\acrodef{ACK}{acknowledgement}
\acrodef{NACK}{not-acknowledgement}
\acrodef{RB}{resource block}
\acrodef{URLLC}{ultra-reliable low-latency communication}
\acrodef{mMTC}{massive machine-type communication}
\acrodef{MU}{multiuser}
\acrodef{DL}{downlink}
\acrodef{UL}{uplink}
\acrodef{SC}{subcarrier}
\acrodef{TTD}{True Time Delay}
\acrodef{QAM}{quadrature amplitude modulation}

\begin{document}
%
\title{Rainbow-link: Beam-Alignment-Free and Grant-Free mmW Multiple Access using True-Time-Delay Array}

\author{Ruifu~Li,~\IEEEmembership{Student~Member,~IEEE},~Han~Yan,~\IEEEmembership{Member,~IEEE},~and~Danijela~Cabric,~\IEEEmembership{Fellow,~IEEE}%
\thanks{Ruifu Li, Han Yan, and Danijela Cabric are with the Electrical and Computer Engineering Department, University of California, Los Angeles, Los Angeles, CA 90095 (e-mail: doanr37@ucla.edu, yhaddint@ucla.edu; danijela@ee.ucla.edu).}
\thanks{This work is supported by NSF under grant 1718742 and 1955672. This work was also supported in part by the ComSenTer and CONIX Research Centers, two of six centers in JUMP, a Semiconductor Research Corporation (SRC) program sponsored by DARPA.}
}




\maketitle

\begin{abstract}
\color{black}
The millimeter-wave (mmW) communications is a key enabling technology in 5G to provide ultra-high throughput. Current mmW technologies rely on analog phased arrays to realize beamforming gain and overcome high path loss. However, due to a limited number of simultaneous beams that can be created with analog/hybrid phased antenna arrays, the overheads of beam training and beam scheduling become a bottleneck for emerging networks that need to support a large number of users and low latency applications. This paper introduces rainbow-link, a novel multiple access protocol, that can achieve low latency and massive connectivity by exploiting wide bandwidth at mmW frequencies and novel analog true-time-delay array architecture with frequency dependent beamforming capability. In the proposed design, the network infrastructure is equipped with the true-time-delay array to simultaneously steer different frequency resource blocks towards distinct directions covering the entire cell sector. Users or devices, equipped with a narrowband receiver and either a single antenna or small phased antenna array, connect to the network based on their angular positions by selecting frequency resources within their rainbow beam allocation. Rainbow-link is combined with a contention-based grant-free access to eliminate the explicit beam training and user scheduling. The proposed design and analysis show that rainbow-link grant-free access is a potential candidate for latency-critical use cases within massive connectivity. Our results show that, given less than $10^{-5}$ probability of packet loss, a rainbow-link cell, over 1 GHz bandwidth using 64 element antenna array, attains sub-millisecond user-plane latency and Mbps user rates with an approximate \SI{400}{m} line-of-sight coverage and a density of up to 5 active single antenna users per second per \si{}{m}$^2$. 
\color{black}
\end{abstract}



%
\IEEEpeerreviewmaketitle

\section{Introduction}
\label{sec:Introduction}

\IEEEPARstart{O}NE objective of the 5G evolution is to further diversify its performance and support applications with improved data rates, latency, and number of connected devices. For example, in the emerging Industry 4.0 \cite{7883994},  it is estimated that communications of \ac{IWSN} in a small cell with up to 1 device per square-meter connection density require 10 to 100 Mbps data rate, with 5 to 30 ms latency, and medium device power consumption and cost \cite{Ericsson_NRlite,QCOM_NRlite}. \color{black} These stringent requirements coincide with all three use categories in 5G specified as \ac{eMBB}, \ac{URLLC} and \ac{mMTC}. Specifically, the \ac{eMBB} is tailored for high peak rate and throughput. \color{black} General requirements for \ac{URLLC} are sub 1ms user plane latency and packet error rate as low as $10^{-5}$ \cite{urllc2014}. For \ac{mMTC}, the general requirement, as suggested by its name, is to provide massive wireless connectivity (e.g. beyond $10^5$ links) to machine-type devices in a given area \cite{massive5gbeyond}.

The \ac{mmW} communications is a key technology in 5G. Due to abundant bandwidth and antenna array beamforming, \ac{mmW} communications is the enabler of \ac{eMBB}. It is also envisioned as a promising candidate for the future high-end \ac{IWSN} \cite{Nokia_NRlite}.
However, the current \ac{mmW} solutions have many disadvantages in terms of latency, connection density, power and cost for the industry \ac{IoT} applications. Firstly, \ac{mmW} systems rely on beamforming to overcome severe propagation loss. To keep reasonable cost and power consumption, current systems utilize phased antenna arrays at both \ac{BS} and \ac{UE} for beamforming. The overhead associated with beam alignment and beam scheduling using analog arrays is a latency bottleneck. Secondly, the analog phased antenna arrays can connect only a limited number of devices based on the number of beams and the number of \ac{RF} chains. Each beamformed transmission occupies the entire bandwidth, therefore frequency domain multiple access cannot be supported. As a result, phased antenna array based \ac{mmW} networks cannot support a high number of connected devices.  Lastly, cost and power consumption of wide-band \ac{mmW} devices are high which presents a major limitation for \ac{IoT} applications. 

\color{black} Meeting diversified service requirements with reasonable cost and power consumption is inherently hard. Specifically, to tackle URLLC requirement, key 5G NR features such as grant-free multiple access must be employed \cite{9521577}. In this regime, the network capacity is often limited by insufficient radio resource elements. \cite{8877253} evaluates URLLC use cases with 10MHz band at 4GHz carrier frequency where the capacity attains 10 users per cell. Alternatively, in \ac{mmW} band large unlicensed band resources are available. However, due to the need of beamforming,  these resources cannot be shared between a large number of spatially separated users \cite{7805314}. To date, most works that focus on latency in \ac{mmW} systems emphasize on reducing the initial access and beam alignment latency \cite{2018FastMW}. Latency due to resource scheduling among multiple users is highlighted by work \cite{7876982}, where optimized frame structure and fully digital arrays were proposed as a solution. The ability of a digital array to connect with users in all directions is useful, however its power consumption and cost is overwhelmingly high. \color{black} Therefore, to address aforementioned weaknesses of existing solutions for \ac{mMTC} and \ac{URLLC}, rethinking of array architectures and multiple access schemes is required.

\begin{figure}
    \centering
    \includegraphics[width = \linewidth]{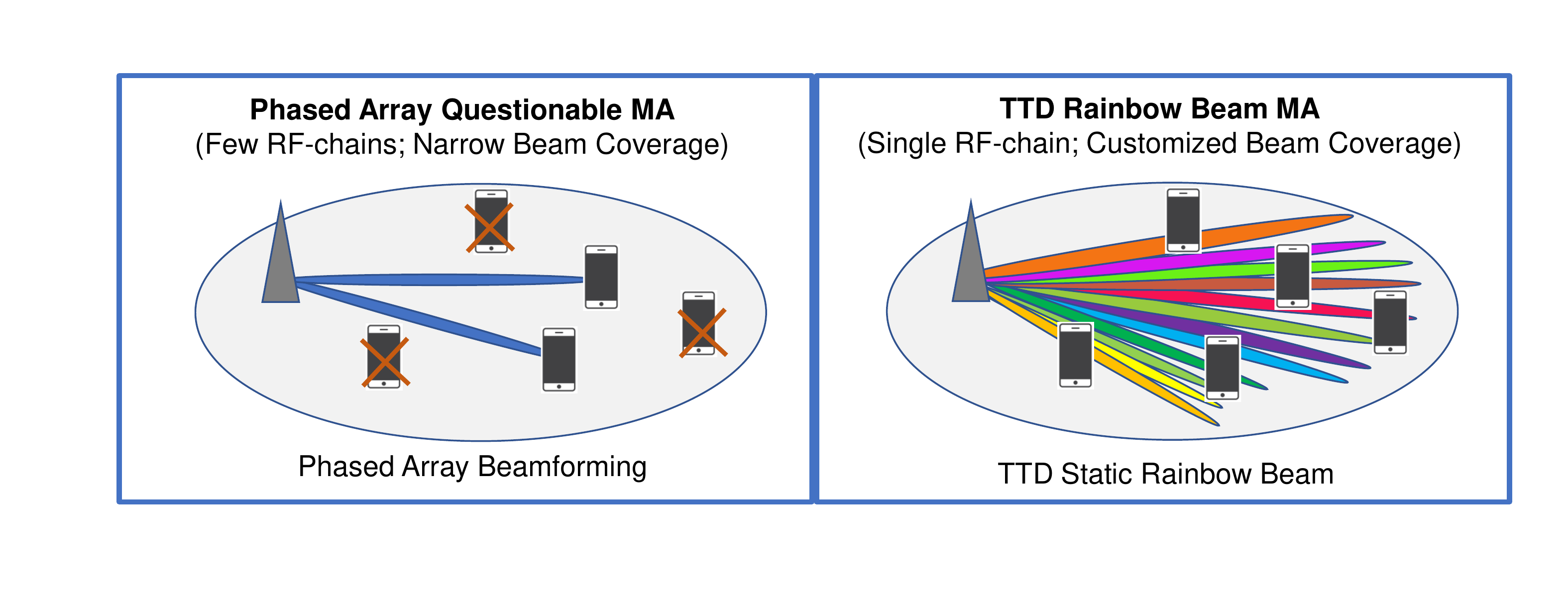}
    \caption{ \color{black} Due to narrow angular coverage of analog beams, \ac{BS} with phased array can only serve a limited number of scheduled users. On the other hand, \ac{TTD} array can connect with users in the entire spatial sector without any scheduling. With phased array, the entire band is pointed to a certain direction, while with \ac{TTD} array, each beam is pointed to a specific frequency subband (represented by colors). \color{black}}
    \label{fig:my_label}
\end{figure}

In this work, we leverage a new beamforming technique referred to as rainbow beam \cite{9048885}, that is enabled by a  novel \ac{TTD} array architecture \cite{ghaderi_ttd}. Rainbow beams exhibit beam patterns that cover the entire angular space by uniquely mapping different frequencies, e.g.  \ac{OFDM} \ac{SC}, onto specific directions. As indicated in \Cref{fig:my_label}, rainbow beams are particularly attractive in wideband systems as different \ac{SC}s and corresponding beamforming directions can be allocated to a large number of users.  
\color{black} As long as the users select their appropriate frequency resources, no beam training or scheduling is needed. Based on this feature, in our design a grant free \ac{mmW} multiple access protocol is combined with rainbow beamforming. It has a potential to provide low user-plane latency for a massive amount of users in \ac{IoT} applications.

The idea of utilizing frequency dependent beamforming to enhance coverage of directional beams (therefore provide multiple access) has been studied in sub-terahertz communication system \cite{9399121}, \cite{leaky_wave}, \cite{delay_phase}.
However, previous studies largely considered broadband, high-rate communications, as opposed to \ac{IoT} use cases.
Differing from most \ac{mmW} networks, in our design the multiple access based on rainbow beams is regarded as an enabling solution for latency-critical \ac{mMTC} \cite{survey_mmtcurllc}. On one hand it provides wide angular beam coverage in a power efficient manner by spatially spreading \ac{SC}s; on the other hand it operates in \ac{mmW} band where sufficient bandwidth resources are available to reduce system overhead. 
\color{black}
\color{black}
Our main contributions are the following:
\begin{itemize}
    \item {We propose a rainbow-beam based frequency domain multiple access which does not require explicit beam-training and supports grant-free random access. 
    }
    \item{We propose a detailed design for radio interface, \ac{DL} synchronization procedure, and random access protocol.}
    \item{We analyze synchronization performance and probability of collision (packet loss) and their impact on latency for given network design parameters.\footnote{
    Our proposed scheme inherently has only two sources of latency due to synchronization and collision with other users that share the same frequency and beam resources.}
    }
    \item{
    Our analytical study is supported by simulations that evaluate the proposed multiple access scheme in terms of latency, reliability and effective rates.
    }
\end{itemize}
\color{black}

The rest of the paper is organized as follows. In \Cref{sec:system_model}, we introduce the system model. The proposed rainbow link protocol and radio interface design are included in \Cref{sec:protocol}. The analysis of \ac{DL} synchronization and \ac{UL} grant-free transmission are presented in \Cref{sec:network_analysis}. Section V presents simulation results. Discussion and future work suggestions are in Section VI. The paper is concluded in Section VII.

\textit{Notations:} Scalars, vectors, and matrices are denoted by non-bold, bold lower-case, and bold upper-case letters, respectively, e.g. $h$, $\mathbf{h}$ and $\mathbf{H}$. The element in $i$-th row and $j$-th column in matrix $\mathbf{H}$ is denoted by $[\mathbf{H}]_{i,j}$. Transpose and Hermitian transpose are denoted by $(.)^{\transpose}$ and $(.)^{\hermitian}$, respectively. The $l_2$-norm of a vector $\mathbf{h}$ is denoted by $||\mathbf{h}||$. $\text{diag}(\mathbf{A})$ aligns diagonal elements of $\mathbf{A}$ into a vector, and $\text{diag}(\mathbf{a})$ aligns vector $\mathbf{a}$ into a diagonal matrix. The $i$-th element in set $\mathcal{S}$ is denoted as $[\mathcal{S}]_i$.

%
%
\section{System Model}
\label{sec:system_model}

\begin{figure*}[htbp!]
\centering
\includegraphics[width=1\textwidth]{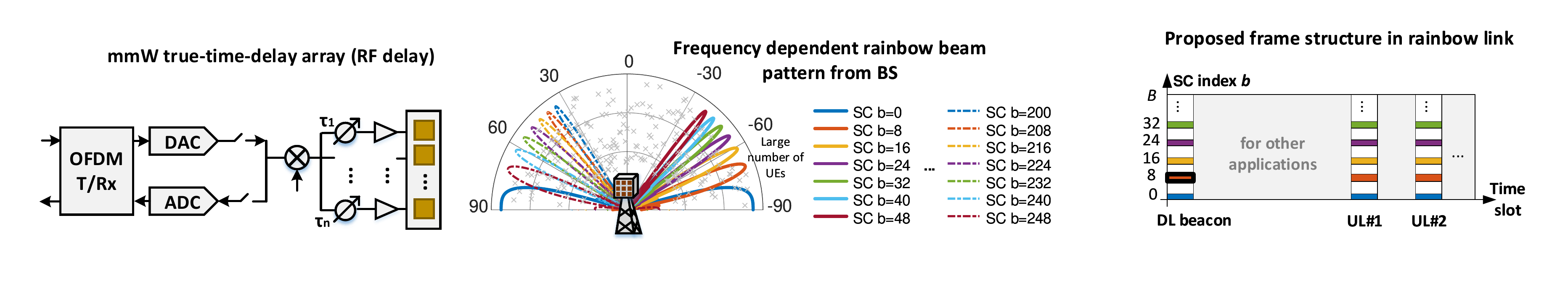}
\vspace{-5mm}
\caption{The left figure shows a typical true-time-delay mmW arrays. The middle figure illustrates the rainbow beam pattern of the base station with $N_{\bs}=64$ antennas. The BS is in origin point of polar plot and gray crosses represent connected user equipment. }
\vspace{-2mm}
\label{fig:rainbow_beam}
\end{figure*}

In this section, we introduce  \color{black} the system model of \ac{mmW} communication with TTD beamforming. \color{black} 
All important notations are summarized in Table~\ref{tab:notation}.

\begin{table} [htbp!]
{\color{black}
\caption{Nomenclature}
\centering
\begin{tabular}{|c|c|}
\hline 
Symbol &Explanations\tabularnewline
\hline 
\hline 
\multicolumn{2}{|c|}{Symbols related to system model}\tabularnewline
\hline 
$b$ & Index of \ac{SC}\tabularnewline
$\mathbf{w}_b$ & beamformer in the BS for $b$-th \ac{SC}\tabularnewline
$u$ & Index of UE\tabularnewline
$\mathbf{v}_u$ & freq independent beamformer at the UE \tabularnewline
$\mathcal{B}_u$ & \ac{SC} set a narrowband UE can access \tabularnewline
$N_{\bs}$, $N_{\ue}$  & Number of antenna in BS and UE\tabularnewline
$\BW$ &  total bandwidth of the network\tabularnewline 
$M$ & Number of multi-path components\tabularnewline
$\mathbf{H}_{u,m}$ & m-th narrowband channel for the u-th UE \tabularnewline
$\mathbf{a}_{\bs}\left(\theta_u\right)$, $\mathbf{a}_{\ue}\left(\phi_u\right)$   & Spatial responses of BS and UE\tabularnewline
$\phi_u,\theta_u$, $g_u$, $\tau_u$  &Path AoA/AoD/gain/delay of $u$-th user  \tabularnewline
\hline
\hline
\multicolumn{2}{|c|}{Symbols related to multiple access}\tabularnewline
\hline 
$n$  & Number of repetitions over resource blocks\tabularnewline
$K$ & Narrowband cardinality of resource blocks\tabularnewline
$B$ & Total number of \ac{SC}\tabularnewline
$U$ & Number of active users per frame\tabularnewline
$p$ & Rate of user activation per frame \tabularnewline
$\ggroup$ & Number of \ac{SC}s bundled into a resource block \tabularnewline
\hline  
\end{tabular}
\label{tab:notation}
}
\end{table}

We consider a \ac{mmW} time-division duplex (TDD) network with center frequency 
$f_c$ and total bandwidth $\BW$. The \ac{OFDM} waveform with \ac{CP} is used with a total number of $B$ \ac{SC}s.

The network consists of a massive number of quasi-static \ac{mmW} \ac{UE}s that are served by a BS. We assume each \ac{UE}, a machine type device, is equipped with a phased array with $N_{\ue}$ antenna. Furthermore, we assume each UE has much narrower bandwidth than the BS to further reduce power and cost. 
As a result, each user can only access a subset of \ac{SC}s through \ac{OFDMA}. The indices of those \ac{SC}s accessible by user $u$ are denoted by set $\mathcal{B}_u$ whose cardinality satisfies $|\mathcal{B}_u| \ll B$, i.e., the \ac{SC}s that a user can access is much smaller than the total number of \ac{SC}s in the broadband.
\color{black}  In practical  \ac{OFDM} based system, adjacent \ac{SC}s are typically bundled into blocks for frequency resource utilization. This mechanism is controlled by $\ggroup$  which denotes how many \ac{SC}s are consolidated into a \ac{RB}. To be clear on notations, we use $|\mathcal{B}_u|$ for the number of \ac{SC}s and $K$ for the number of \ac{RB}s in $\mathcal{B}_u$, i.e., $K = \left|{B}_u\right|/\ggroup$
\color{black}.

We assume the \ac{BS} has a linear array with $N_{\bs}$ elements. The antennas are critically spaced, i.e., half of wavelength that associates with $f_c$. In this work we focus on a sparse  geometric channel of $M$ multipath components\footnote{For mmW communication, typically 
$M \leq 4$}, where

the \ac{AoA}, \ac{AoD}, and complex path gain of the $u$-th \ac{UE} and $m$-th multipath component are denoted as $\theta_{u,m}$, $\phi_{u,m}$, and $g_{u,m}$, respectively.
$\theta_{u,m}$ and $\phi_{u,m}$ are assumed to be uniform randomly distributed in region $-\pi/2$ to $\pi/2$. 
The channel $\mathbf{H}_{u,b}$, for $u$-th user and $b$-th \ac{SC} is
\begin{align}\label{eq:channel}
    \mathbf{H}_{u,b} =\frac{1}{\sqrt{N_{\bs}N_{U}}}\sum\limits_{m = 1}^M \alpha_{u,b,m}\mathbf{a}_{\ue}(\phi_{u,m})\mathbf{a}^{\hermitian}_{\bs}(\theta_{u,m})
\end{align}
where $\alpha_{u,b,m} =  g_{u,m}\mathrm{exp}(j2\pi b \tau_{u,m} \BW/B)$ is the \ac{SC}-wise complex gain, with $g_{u,m}$ and $\tau_{u,m}$ representing the multipath gain and delay with respect to the first antenna element as a reference. We denote $\mathbf{a}_{\bs}(\phi)$ and $\mathbf{a}_{\ue}(\theta)$ as the narrowband\footnote{The narrowband array response model holds true when the propagation delay across the array aperture is less than the sampling duration \cite{9048885}. For typical \ac{mmW} BS array aperture $\leq 0.32$m, the propagation delay across the aperture is up to 1 ns, which is less than 2.5ns, the sampling duration of a 400MHz system.} array response vectors, i.e.,$    [\mathbf{a}_{\ue}(\theta)]_n =\mathrm{exp}\left[j\pi (n-1)\sin(\theta)\right]$ and $[\mathbf{a}_{\bs}(\phi)]_n =\mathrm{exp}\left[j\pi (n-1)\sin(\phi)\right]$. Amplitude of 
$g_{u,m}$ is characterized by the free space path loss model \cite{6824746}. 

In general, channel with multiple taps would be frequency selective. However, since all \ac{UE}s operate over a small portion of broadband bandwidth, we assume the channel is approximately flat. 
 We also assume the \ac{BS} is equipped with a reconfigurable \ac{TTD} array with a single \ac{RF} chain for its transceiver as shown in Figure~\ref{fig:rainbow_beam}. 
In our previous work, we have shown that TTD array can realize frequency dependent beams so that each \ac{OFDM} \ac{SC} is mapped to a particular beamforming direction \cite{9048885}. Delay taps in the array are set to be uniformly spaced with inter-element delay spacing of $\Delta \tau$, regardless of whether the delays are introduced in baseband or RF. The \textit{full-range} rainbow beam\footnote{It is noted that this rainbow beam operation has strict requirement on the delay range of the circuit blocks that introduce delay. However, with the increased bandwidth in \ac{mmW} and sub-THz, as well as the recent break-through of circuits \cite{lin20214element}, such regime is attainable.} can be achieved with tap $\Delta\tau = 1/\BW$ such that the equivalent analog combiner for the $b$-th \ac{SC} is given by
\begin{align}
    [\mathbf{w}_{b}]_i = \mathrm{exp}\left[j2\pi b (i-1)/B\right].
\label{eq:rainbow_beam_steering_vector}
\end{align}
Inspecting (\ref{eq:rainbow_beam_steering_vector}), it is clear that beamformer $\mathbf{w}_{b}$ steers \ac{SC} $b$ onto a unique directions with respect to the rest of \ac{SC}s. 
On the other hand, any steering vector, i.e., $\mathbf{a}_{\bs}(\phi)$ for an arbitrary $\phi$, is also synthesized at a certain \ac{SC}. Hence it is referred to as full coverage rainbow beamforming. \color{black} For a certain user, we define $\bstaru$ as the index of \ac{SC} whose encoded spatial direction is the closest to the \ac{AoA} $\theta_{u}$ of a certain path. This particular \ac{SC} is also denoted as \textit{anchor \ac{SC}}. In other words, the beamforming gain at anchor \ac{SC} is the highest among all subcarriers. \color{black}

To better present our proposed network protocol, we briefly introduce the signal model in the \ac{DL} and \ac{UL}. In the DL, we denote the transmit \ac{OFDM} symbol at the $b$-th \ac{SC} as $S^{(\text{DL})}[b]$. Given an ideal bandlimited filter at the UE defined by its narrowband $\mathcal{B}_u$, the received frequency domain signal is 
\begin{align}
    R_u^{(\text{DL})}[b] =
    \underbrace{ \mathbf{v}_u^{\hermitian}\mathbf{H}_{u,b}\mathbf{w}_{b}}_{\beta_{u,b}}S^{(\text{DL})}[b] + z^{(\text{DL})}[b], \text{for} ~ b\in\mathcal{B}_u.
\label{eq:freq_DL_symbol}
\end{align}
 In the above equation, $\mathbf{v}_u$ represents the precoder of the $u$-th user (which is not frequency dependent) and scalar $\beta_{u,b}$ characterizes the post-beamforming gain of the $b$-th \ac{SC}. $z^{(\text{DL})}[b]$ refers to the noise at the $b$-th \ac{SC}. 
.

The received signal at the BS in the \ac{UL} is expressed as
\begin{align}
R^{(\text{UL})}[b] = \sum_{u=1}^{U}\underbrace{\mathbf{w}_{b}^{\hermitian}\mathbf{H}^{\hermitian}_{u,b}\mathbf{v}_u}_{\beta^{*}_{u,b}}S^{(\text{UL})}_{u}[b] + z^{(\text{UL})}[b]
\label{eq:freq_UL_symbol}
\end{align}
where the post-beam channel gain in the \ac{UL} $\beta^{*}_{u,b}$ is the conjugate of the one in the \ac{DL} as long as the same \ac{SC} and UE-side beamformer $\mathbf{v}_u$ is used. 

Since each \ac{SC} is mapped to a spatial direction, UE can measure the receive power on \ac{SC}s during \ac{DL} boardcasting and identify the segment of \ac{SC}s with the highest received power as the segment aligned with its \ac{AoA}. Note that although the channel gain is assumed to be flat across the narrowband $\mathcal{B}_u $, post-beamforming gain
$\beta_{u,b}$ is still frequency dependent ($b$ dependent) due to TTD beamforming precoder $\mathbf{w}_b$ at the BS side. \color{black} This unique feature of rainbow-beam with TTD array causes beamforming gain loss on different \ac{SC}s. Even if a \ac{UE} has broadband sampling capability, there will only be certain \ac{SC}s that it can use. This design consideration is addressed in Section \ref{sec:network_analysis} \color{black}.

%
%
\section{Rainbow Link Medium Access Control}
\label{sec:protocol}


In a conventional \ac{mmW} \ac{UL} access, a 4-step random access procedure is employed. The \ac{BS} broadcasts synchronization and beam training pilots during \ac{DL}. The UEs independently conduct synchronization, measure the received signal strength of BS's beams, and send feedback through the random access channel. After the \ac{BS} receives feedback, it applies user scheduling and resource allocation scheme by reserving dedicated time-frequency resources for UEs before sending them access grant. \ac{UE}s then use the scheduled resources to complete \ac{UL} transmission. In this scheme a non-trivial latency is expected not only in the grant request, but also in the resource scheduling because the served UEs in each time slot are limited to a small angular region covered by a narrow analog beam from the \ac{BS}.

The rainbow link on the other hand, leverages the frequency-dependent beam steering capability of \ac{TTD} array which can simultaneously connect multiple UEs. We propose to use a fixed beam configuration given by beamforming precoder/combiner in (\ref{eq:rainbow_beam_steering_vector}) for both \ac{DL} synchronization and \ac{UL} access, i.e., on the \ac{BS}'s side no beam switching is required. Due to large coverage of rainbow beam in angular domain, \ac{UE}s always have beamforming gain to and from \ac{BS} without explicit beamforming training or feedback. 
\color{black}
Elimination of beamforming feedback significantly cuts the overhead on beam maintenance and resource scheduling which then makes the 2-step contention based random access possible.  
\color{black}

This section presents the proposed protocol for multiple access with rainbow link. We start by discussing details of \ac{DL} synchronization and \ac{UL} grant-free transmission. Based on the discussion, we then provide the design of radio interface as well as relevant performance metrics that are analyzed in Sections IV and V.

\subsection{Rainbow beam DL synchronization}
\label{subsec:protocolsync}

Given a wideband  \ac{OFDM} transmission from the \ac{BS}, the synchronization of the narrowband UEs require estimation of the correct timing offset and \ac{FFT} window for decoding. Effectively, a narrowband receiver at the \ac{UE} side needs to select a set of \ac{SC}s $\mathcal{B}_u$ that are then used for \ac{UL} transmission. Since rainbow beam maps each \ac{SC} to a specific spatial direction, \ac{UE}s should select \ac{SC}s that are mapped to its \ac{AoD} so that they leverage the maximum beamforming gain from \ac{BS} and improve \ac{DL} for the detection of the synchronization signal. Due to narrowband receiver processing at the \ac{UE}, the synchronization involves following challenges:
\begin{itemize}
    \item {\ac{UE}s should synchronize using a received signal that contains only a fraction of the entire OFDM preamble.  }
    \item {\ac{UE}s needs to locate (in the frequency domain) a segment of \ac{SC}s  with the minimum beamforming gain loss as its operating band $\mathcal{B}_u$.}
\end{itemize}

To tackle the first challenge, we propose to load synchronization sequences on all SCs.
Since the  \ac{AoD} of a \ac{UE} in \ac{DL} channel might land on any \ac{SC} across the wideband, we employ \ac{PN} sequences instead of Zadoff-Chu sequences that are used in 4G \ac{NB-IoT}. \ac{PN} sequences have a low peak to average power ratio, zero auto-correlation with its time-shifted version, and do not require fixed narrowband reception for completeness \cite{8055639}.  The preamble sequence is assumed to be known at \ac{UE}s. 

The second challenge is essentially dependent on the UE's ability to maintain beam alignment with the \ac{BS}. With no prior information about which segment it should select, a UE would have to traverse the entire broadband signal and repeatedly search for $\bstaru$, the anchor \ac{SC} with the highest beamforming gain. Undoubtedly, traversing the entire broadband frequencies (with low sampling rate) would cause prohibitive overhead. We assume that \ac{UE}s are quasi-static and thus the \ac{AoD}s change slowly. In this case, using the previously selected segment as a prior, \ac{UE}s can keep track of changes in AoDs. This simplification eliminates the extra overhead of \ac{UE}s searching for $\bstaru$. 

 As illustrated in \Cref{fig:rainbow_beam}, synchronization sequences are loaded onto different \ac{SC}s in the \ac{DL} broadcast. The \ac{DL} synchronization signal contains frequency multiplexed narrowband beacons\footnote{This concept is similar to NB-IoT where a narrowband synchronization sequence is loaded in resource block. What is unique in rainbow link is that each anchor carrier is coded with a unique analog beam.}. In order to participate in contention UEs are required to successfully synchronize.  The synchronization algorithm at the UE involves the following steps:
\begin{itemize}
    \item {Perform down conversion by multiplying the broadband signal with $ \exp\left(-1\mathrm{j}2\pi t \bstaru/B \right)$, filtering, and sampling to obtain a narrowband baseband signal. $\bstaru$ here refers to the anchor \ac{SC} from the previous connection.}
    \item {Locate the correct \ac{FFT} window by estimating the integer and fractional timing offset.}
    \item {Identify the new anchor \ac{SC} $\bstaru$ with the highest signal strength. The updated $\mathcal{B}_u$ for \ac{UL} transmission is determined accordingly.}
\end{itemize}
\color{red}

\color{black}

\subsection{Rainbow Beam \ac{UL} Grant-free Transmission}
\label{subsec:protocolcolli}

\begin{figure}
\centering
\begin{tabular}[b]{c}
    \includegraphics[width=0.90\linewidth]{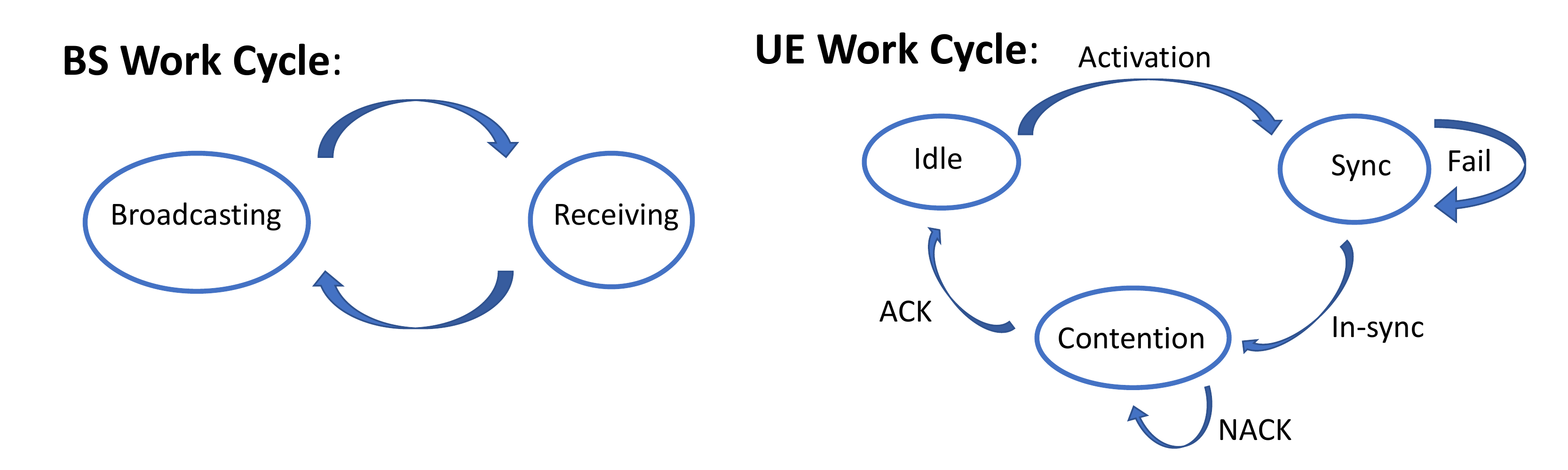}  \\
      \small (a) BS \& UE Work Cycles \\
    \includegraphics[width=0.90\linewidth]{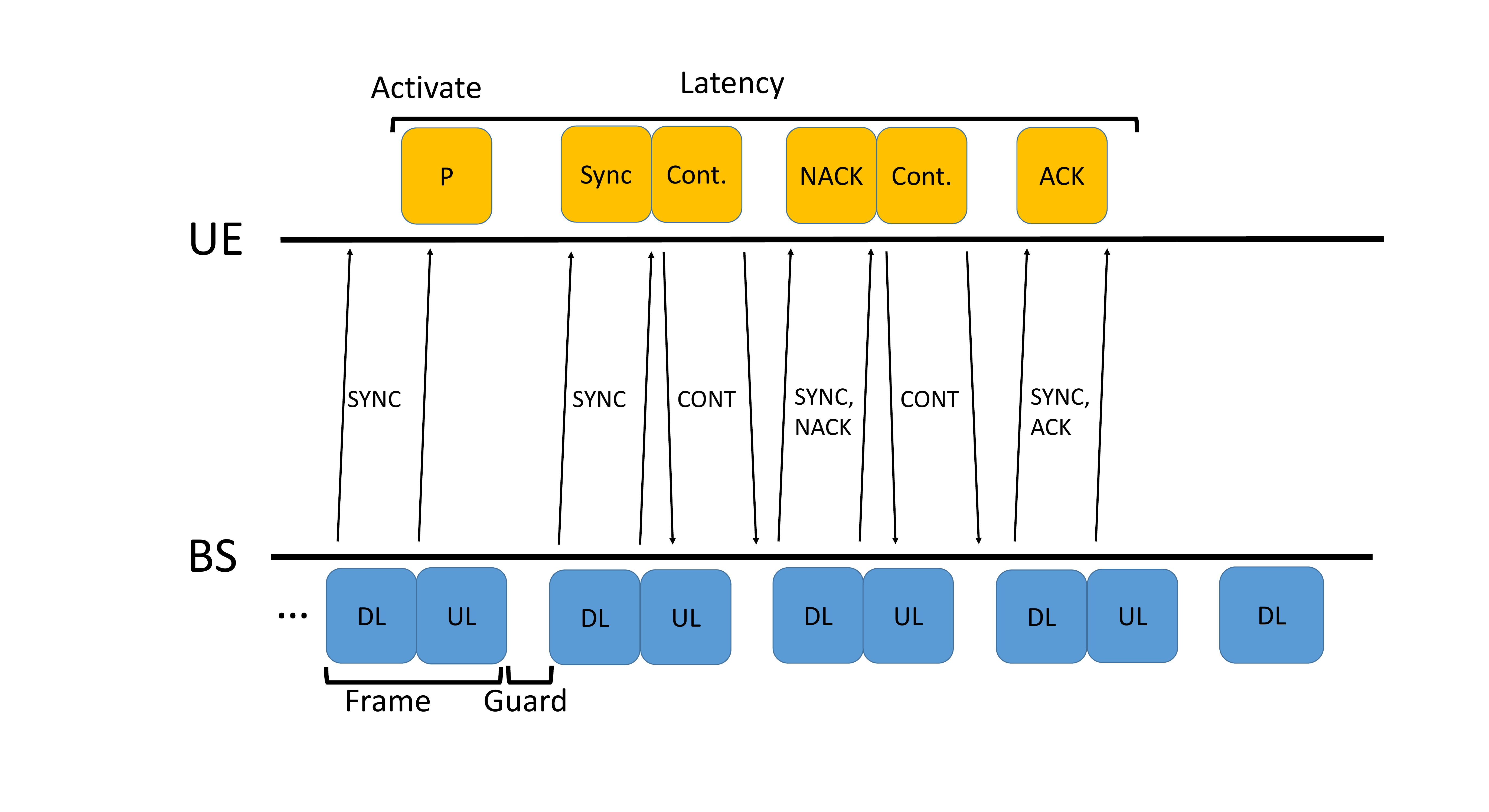}  \\
      \small (b) User Plane Latency \\
\end{tabular}
\caption{The uplink grant-free access with proactive repetition scheme similar to \cite{8269137,yliuetal}. The abbreviation \textbf{P}, \textbf{Cont}, \textbf{ACK}, and \textbf{Sync} denote packet arrival, contention attempts, \ac{ACK} and synchronization attempts, respectively. A short guard period counts for propagation and processing delay. In the diagram, \ac{UE} succeeded in the first synchronization attempt and the second contention attempt}
\vspace{-1mm}
\label{fig:lat_concept}
\end{figure}

The proposed work cycles and the corresponding timing diagram of the \ac{UL} grant-free protocol is shown in \Cref{fig:lat_concept}. When their packets are generated, each \ac{UE} activates and synchronizes in the nearest \ac{DL} time frame where \ac{BS} broadcasts synchronization signal, after which \ac{UE} transmits in an arrive-and-go manner. Namely, \ac{UE} does not send a scheduling request for access grant as in grant-based access. For the \ac{UL} transmission it utilizes multiple \ac{RB}s in the narrow band centered\footnote{From a geometric argument, putting the anchor \ac{SC} in the center of $\mathcal{B}_u$ intuitively ensures that there is good channel gain.}
at the detected anchor \ac{SC}, i.e., $\mathcal{B}_u = [\bstaru-\left|\mathcal{B}_u \right|/2, \bstaru+\left|\mathcal{B}_u \right|/2 - 1]$. UEs are sending packets that contain preamble and data so that BS can extract UE's identity and other necessary control information.


Naturally in a grant-free access, multiple packets transmitted in the same time-frequency resource block from different \ac{UE}s would cause a collision and packet loss. For the sake of tractable analysis, we assume decoding on \ac{SC}s without collision always succeeds, and decoding on \ac{SC}s with collision always fails. In order to improve the packet loss rate, we propose a simple repetition coding. With abundant time-frequency resource blocks in \ac{mmW} spectrum, \ac{UE}s can perform repetitive transmissions on multiple resource blocks \cite{Contention_Based}. 
In our proposed scheme, each \ac{UE} repeats its transmission on $n$ \ac{RB}s in its narrow band of $K$ \ac{RB}s. The $n$ replicas offer repetition coding and diversity gain to reduce packet loss. 
\color{black} Making a large number of repetitions alleviates the loss of packets for a certain user. But the contention also becomes harsh when other UEs do so. \color{black} In this work we assume a fixed number of repetition $n$ for all users and study the optimal design of $n$. Notice that the optimal $n$ will also be influenced by the notion of grouping. When multiple \ac{SC}s are consolidated as a \ac{RB}, \ac{UE}s load more useful symbols into a packet but have less resources to avoid collisions. We will analyze these trade offs in Section IV. 

\subsection{Radio Interface Design}

In order to analyze rates, coverage, and latency of the proposed system, this section briefly discusses radio interface design. Due to its popularity, reliability, and flexibility for contention based grant-free random access \cite{yliuetal}, we adopt a frame slotted Aloha. Both \ac{UL} and \ac{DL} transmissions are slotted in terms of frames\footnote{The propose protocol utilizes subset of time-frequency resource, similar to the interleaved NB-IoT resource units with LTE \cite{NB_IOT2017}, and we focus on relevant UL/DL only.} and users can transmit without grant request or allocated resources from \ac{BS}. Critical specifications in radio interface for the proposed system include \cite{7063633}:

\begin{itemize}
    \item {Subcarrier spacing and  \ac{OFDM} symbol duration.} 
    \item {Cyclic prefix to accommodate combined channel delay spread and the largest delay tap of TTD array} 
    \item {Number of \ac{UL} \& \ac{DL} TTIs}
    \item {Length of \ac{UL} \& \ac{DL} preamble}
    \item {Payload in \ac{DL} frames}
\end{itemize}

In \ac{URLLC} and \ac{mMTC} use cases, short packets are commonly used \cite{8403963,7529226}. In that sense, large spacing between \ac{SC}s shall be considered. A natural choice is then to adopt a numerology scaled from the current 5G and 4G networks. As pointed out by \cite{8412469}, in current standards, \ac{SC} separation (SCS) is of the form: $2^k \times 15 \mathrm{kHz}$ for integer $k$. Specifications with $k > 2$ cannot be implemented in sub-6GHz spectrum. To get a rough understanding of the system, we then use the $k = 5$ case as listed in \Cref{tab:numer1}. Here the length of \ac{CP} is about $250$ns which is sufficient to accommodate the combination of largest delay of \ac{TTD} element\footnote{The largest delay is $N_{\mathrm{B}}$ samples as specified by size of the array. $256$ samples could accommodate a $128$ element array.} and the root mean square delay spread of \ac{mmW} frequencies in most outdoor environment \cite{6824746}. 

Each frame would contain multiple \ac{TTI}s.  Frame length is a critical design to accommodate various control data and payloads. 
With the grant-free multiple access, the structure of  \ac{DL} \ac{TTI} in the proposed scheme is relatively simple. The detailed frame structure is specified in \Cref{tab:linkbud}.

\begin{table} [htbp!]
\caption{Numerology \color{red}\color{black}}
\centering
\begin{tabular}{|c|c|}
\hline 
Specification & Value\tabularnewline
\hline 
\hline 
Sampling Frequency[MHz] & 983.04\tabularnewline
Number of Points in FFT  & 2048\tabularnewline
$\BW$ [MHz]  & 1000\tabularnewline
\hline  
\ac{SC} Spacing [kHz] & 480\tabularnewline
Cyclic Prefix [Samples] & 256 \tabularnewline
Symbols per TTI & 13 \tabularnewline
Symbol Duration [$\mu s$] & 2.083\tabularnewline
Symbol Duration with $CP$ [$\mu s$] & 2.344\tabularnewline
TTI Duration [$\mu s$] & 30.469\tabularnewline
\hline 
\end{tabular}
\label{tab:numer1}
\end{table}

Although \ac{BS} transmits wideband signals, only a portion of the band is accessible to each narrowband \ac{UE}. Therefore both \ac{UL} and \ac{DL} reference signals (preambles) in our design are much longer than the ones in \cite{7063633} due to lack of broadband receiving capabilities. We also assume open-loop power control with a maximum UE transmit power of  $23$ dBm \cite{5198853} which allows for shorter \ac{UL} reference 
symbols than that of \ac{DL} frame.  

\begin{table} [htbp!]
\caption{Frame Design}
\centering
\begin{tabular}{|c|c|}
\hline 
Specification &  Design Values \tabularnewline
\hline 
\hline 
UL Symbols [TTI] & 2 \tabularnewline
DL Symbols [TTI] & 2 \tabularnewline
Frame Length [$\mu s$] & 125\tabularnewline 
Frame Length [symbols] & 52 \tabularnewline 
\hline 
DL Reference Signal [symbol] & 22 \tabularnewline
DL Control Signal [symbol] & 2  \tabularnewline
Guard Interval [symbol] & 2  \tabularnewline
UL Preamble [symbol] & 8 \tabularnewline
UL Payloads [symbol] & 18 \tabularnewline
\hline 
\end{tabular}
\label{tab:linkbud}
\end{table}

The proposed numerology and frame design are used throughout the rest of the paper. It is worth mentioning that this numerology is tailored to achieve \ac{URLLC} with large \ac{SC} spacing and mini-slots. In general, depending on the application, there are designs with more efficient control fields.  However, these investigations are beyond the scope of this paper.   

\subsection{Performance Metrics}

 To simplify the discussion, we assume that a successful \ac{DL} synchronization 
    and \ac{UL} transmission without collision together lead to a successful packet delivery. During each \ac{DL} broadcasting, \ac{BS} encodes feedback per \ac{SC} to indicate whether previous \ac{UL} payload on that \ac{SC} is successfully decoded. Synchronized \ac{UE}s would persistently transmit until their packets are delivered, while \ac{UE}s with unsuccessful synchronizations do not participate in contention. 
   The probability of successful synchronization is dependent on the receiver \ac{DL} preamble. For users that are within the rainbow link coverage, we analyze the following performance metrics:

\begin{itemize}
    \item { Access latency:  defined as the amount of time between packet arrival and its successful delivery. }
    \item { Effective rate: defined as the ratio of the number of data symbols (payload) in a packet and the time required to successfully delivery through contention.}
\end{itemize}

Both metrics depend on several factors including SNR, user density, the number of repetitions, and grouping of subcarriers into \ac{RB}s. 

The proposed system is expected to serve a large number of spatially separated users with low latency. In the next section, we analyze \ac{DL} synchronization and \ac{UL} contention, and mathematically formulate performance metrics.

\section{Performance Analysis of Rainbow Link}
\label{sec:network_analysis}

\begin{figure}[htbp!]
     \centering
     \includegraphics[height = 0.64\linewidth]{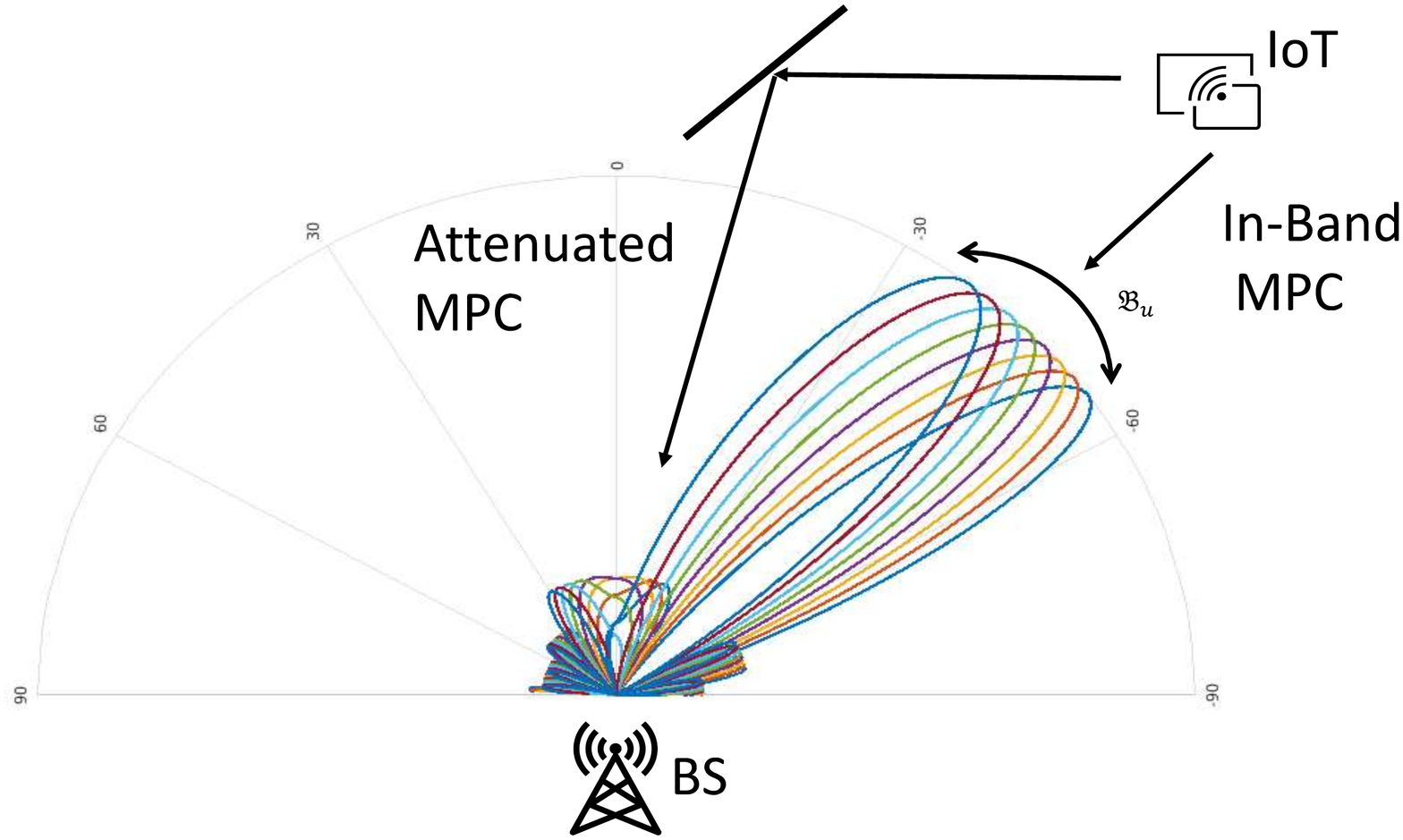}
     \caption{The narrowband operation of UE and rainbow beams naturally reject multipath components. For signal at a given subcarrier, its multi-path components propagate via a different \ac{AoD} and hence
     are filtered out by \ac{UE} due to its limited bandwidth.}
     \label{fig:mpcinband}
 \end{figure}

 In this section, we provide analysis of the proposed rainbow link access. For the sake of tractable analysis and concise notation, we make two additional assumptions. Firstly, we mainly focus on a single path channel, \ie, the summation and the $m$-dependency 
 in (\ref{eq:channel}) are dropped. The underlying rationale is that each \ac{UE} can only access a narrow bandwidth $\mathcal{B}_u$, therefore it is very likely that only a single path aligns in the spatial direction encoded on \ac{SC}s within $\mathcal{B}_u$ as illustrated in \Cref{fig:mpcinband}. Secondly, we omit the discussion of beamforming in the UE's side.  Specifically, $N_{\ue} = 1$ is used, and we drop beamformer $\mathbf{v}$ and array response $\mathbf{a}_\ue \left(\phi_u\right)$ of UE in the analysis. The rationale behind is that in the term $\beta_{u,b} = \mathbf{v}_u^{\hermitian}\mathbf{H}_{u,b}\mathbf{w}_{b}$, \ac{UE}'s beamformer  serves merely as another factor that has no dependency on $b$. In other words, a multi-antenna \ac{UE} results merely in a higher beamforming gain and is not related to any unique feature of rainbow beamforming\footnote{Presumably, using multiple antennas on UE side also requires extra overhead for beamforming training.}.
 
\subsection{Analysis of \ac{DL} synchronization}

\color{black}
Synchronization based on \ac{OFDM} waveform usually involves estimation of timing offset and carrier frequency offset \cite{7569029}.  With TTD array, in addition to timing offset \ac{UE}s need to locate their anchor \ac{SC}s. Here we employ a similar analysis as in \cite{7569029} of \ac{NB-IoT} without addressing the carrier frequency offset.
\color{black}
In order to locate the anchor \ac{SC} $\bstaru$ with the highest gain during \ac{DL} synchronization, a user needs to find:
\begin{align}\label{eq:anchorchoice}
\bstaru = \mathrm{argmax}_{b} \left|\beta_{u,b}\right|^2
\end{align}
$\bstaru$ is then the anchor \ac{SC} whose steering angle is the most aligned with $\theta_u$. Plugging (\ref{eq:channel}), (\ref{eq:rainbow_beam_steering_vector}) into (\ref{eq:freq_DL_symbol}) and simplifying:
\begin{align}
\begin{split}
\left|\beta_{u,b}\right|^2 
=  \frac{|\alpha_u|^2}{N_\bs N_\ue}
\underbrace{\left|
    \frac{\sin \bigg[\frac{N_\bs\pi}{2}\left(\frac{2b}{B}- \sin\theta_u\right)\bigg]}{
    \sin \bigg[\frac{\pi}{2}\left(\frac{2b}{B}- \sin\theta_u\right) \bigg]}\right|^2}_{F_{N_\bs}\left(\pi\zeta/B\right)}\\
\end{split}
\label{eq:rainbow_beam_gain}
\end{align}

The expression above leads to Fejér kernel $ F_{N_\bs}(\cdot)$ which is defined as $F_{N}(x) = \left|\sin\left(Nx/2\right)/\sin\left(x/2\right)\right|^2$. $\bstaru$ should then satisfy the following criterion:
\begin{center}
    $-1 \leq 2\bstaru - B\sin\theta_u \leq 1$
\end{center}

From (\ref{eq:rainbow_beam_gain}), one can see that as the total number of \ac{SC}s in UE's band increases, the edge \ac{SC} $\bstaru \pm \left|\mathcal{B}_u\right|/2$ have an increasing gain loss compared with the anchor \ac{SC} $\bstaru$. This has two implications. Firstly, to ensure a relatively flat beamforming gain across the UE narrowband bandwidth, the total number of \ac{SC}s in the broadband should be much larger than number of antennas at the BS $N_\bs$, \ie, $N_\bs/B \ll 1$. This is because the gain difference in narrow band is only relevant to the ratio $N_\bs/B$ as shown in  (\ref{eq:rainbow_beam_gain}). \color{black} Secondly, due to the gain loss, there are only limited number of \ac{SC}s that can be included
in $\mathcal{B}_u$. For instance, a generic threshold $ \left|\mathcal{B}_u\right| \leq B/N_\bs$ ensures that the \ac{SC} at the edge of $\mathcal{B}_u$ has less than 3dB gain loss compared with the anchor \ac{SC} \color{black}.
Increasing number of antennas $N_\bs$ at the \ac{BS} might result in large gain difference which then limits \ac{SC}s that a user can access. Notice that these phase and gain differences do not affect \ac{UL} data transmission since they can be treated as part of the channel. With \ac{UL} preamble sequence, \ac{BS} estimates the channel per \ac{SC} and thus naturally compensates it.

\color{black}The rest of the subsection then addresses the estimation of timing offset. \color{black} Let $L$ denote the number of symbols in \ac{DL} preamble sequence. Let $l$ be the index of samples in a  \ac{OFDM} symbol such that $ -N_{CP} \leq l \leq |\mathcal{B}_u|$. Then $l$-th sample of the $m$-th symbol in preamble sequence can be expressed as: 
\begin{equation}\label{eq:td_symbol}
\begin{split}
 s^{(\text{DL})}_u[l;m] & = \sum\limits_{b \in \mathcal{B}_u} g_u\left[\mathbf{a}_{\bs}^\hermitian\left(\theta_u\right)\mathbf{w}_b\right]S_u^{(\text{DL})}[b;m]e^{1\mathrm{j}2\pi \cdot \frac{b}{\left|\mathcal{B}_u\right|}\left(l - D_u\right)}  \\
  & +z[l;m] \\ 
\end{split}
\end{equation}
where $D_u$ amounts to normalized timing offset that includes an integer and a fractional part due to down sampling. In the equation above, only \ac{SC}s within the narrow band  $\mathcal{B}_u$ are taken into account. This assumes that \ac{UE}s sample the wideband signal without aliasing. In practice, such assumption can be achieved with a high quality low pass filter. 
With channel gain $g_u$ remaining the same for $ \mathcal{B}_u$ during $L$ symbols, the received frequency domain symbol is then:
\begin{equation}\label{eq:freq_sym}
\begin{split}
  \tilde{S}^{(\text{DL})}_u[b;m] & = g_u\left[\mathbf{a}_\bs^\hermitian\left(\theta_u\right)\mathbf{w}_b\right]S_u^{(\text{DL})}[b;m]e^{1\mathrm{j}2\pi \cdot \frac{b}{\left|\mathcal{B}_u\right|}\left(- D_u\right)}  \\
  & +\tilde{z}[b;m] \\ 
\end{split}
\end{equation}
In (\ref{eq:freq_sym}), $\tilde{z}$ is the per \ac{SC} noise in frequency domain. Synchronization is essentially an estimation of $D_u$ from observed samples $\tilde{S}^{(\text{DL})}_u[b;m]$. However, since $\theta_u$ is unknown, 
\color{black} the gain loss cannot be accurately modeled and causes degradation in synchronization performances. We evaluate the following estimator for timing offset and its synchronization performance: 
\color{black}
\begin{align}\label{eq:syncUE}
  \begin{split}
       D_u^* & = \mathrm{arg} \max\limits_{D_u} 
      \quad \left|J_u\left(D_u\right)\right|^2 \\
      J_u \left(D_u\right) & = \sum\limits_{m,b} 
     \left(\tilde{S}^{(\text{DL})}_u[b;m]\right)^* S_u^{(\text{DL})}[b;m]
       \times e^{1\mathrm{j}2\pi \cdot \frac{b}{\left|\mathcal{B}_u\right|}D_u}
      \\
  \end{split}
\end{align}

\color{black}In (\ref{eq:syncUE}) \ac{UE}s merely perform correlation between $\tilde{S}^{(\text{DL})}_u[b;m]$ 
$S_u^{(\text{DL})}[b;m]$ when estimating $D^*_u$. As $\left|\mathcal{B}_u\right| $ increases, the summation in  (\ref{eq:syncUE}) involves more symbols, but the correlation between the two sequences is also more distorted. 
 To conclude,  in \ac{DL} synchronization \ac{UE}s suffer from gain loss due to the rainbow beam pattern and can therefore only operate on a limited number of \ac{SC}s. This effect will be analyzed via numerical simulations. 
\color{black}

We note that as a unique feature of frequency dependent rainbow beam pattern, the synchronization does benefit from beamforming gain \color{black} contributed by \ac{BS} antenna.
\color{black}The length of preamble $L$ should be designed accordingly to fulfill high probability of detection and low synchronization error.

\subsection{Analysis of packet loss rate}

Next we analyze collisions and \ac{PLR} for given radio resources and user density. The purpose of the analysis is to mathematically guide the design of various system parameters. 

Since UEs are narrowband, there is no collision between \ac{UE}s that contend in non-overlapped bands. \color{black} Given that all users choose a consecutive $K$ \ac{RB}s, \color{black} a subband
would suffer from collisions from users landing on a total number of $2(K - 1)+ 1 = 2K - 1$ \ac{RB}s. To tackle all the possibilities for a given number of repetitions, a combinatorial approach is needed. In the analysis, we focus on a specific user occupying bandwidth $ \mathcal{B}_u$ and calculate the density function of the number of occupied \ac{SC}s in the band before this user adds its replicas. From there, packet loss rate is simply the probability that all replicas from this user are covered by already occupied \ac{SC}s.  
 \color{black} Based on our model, the packet loss rate $\Pplr \left(B,K,U,n,\ggroup \right)$ is given in the following proposition as a function of the number of \ac{SC}s $B$, the number of active users $U$, the number of \ac{RB}s that a user can access $K$, the grouping factor $\ggroup$, and the number of repetitions $n$ for each user\color{black}.
\begin{proposition}
With a uniform spatial distribution of users, an approximated expression of \ac{PLR} of the proposed rainbow link is
\begin{equation}
\label{eq:PLRraw}
\begin{split}
    \Pplr  = \sum\limits_{i = 1}^{U - 1} f\left(U - 1, \ggroup \frac{2K-1}{B},i\right)
     \sum\limits_{ j = n}^{K} \left[\mathbf{T}^i\mathbf{p}_0\right]_j \frac{\mathrm{C}_{j}^n}{\mathrm{C}_{K}^n}, \\
\end{split}
\end{equation}
where $f\left(x, p, i\right) = \mathrm{C}_x^i\left(1 - p\right)^{x - i}p^i$ is the binomial probability distribution function, $\mathrm{C}_x^i$ is the operator that evaluates $x$ choose $i$, $\mathbf{p}_0 = [1,0,\cdots,0]^\transpose \in \mathbb{R}^{K+1}$ is the unit vector, and \color{black} the entry in $m_1$-th row and $m_2$-th column of the matrix $\mathbf{T} \in \mathbb{R}^{ (K+1)\times (K+1)}$ is given by \color{black}
\begin{align}
\begin{split}      
        \left[\mathbf{T}\right]_{m_1,m_2} & =
        \frac{
       \mathrm{C}_{K - m_1}^{m_2 - m_1}
        \mathrm{C}_{m_1}^{n - (m_2 - m_1)}}{(2K-1)\mathrm{C}_{K}^n}\\
        & +\sum\limits_{j = 0}^n\sum\limits_{k = 2}^{K}
     \frac{2\mathrm{C}_{K - k}^{n - j}
       \mathrm{C}_{K - m_1}^{m_2 - m_1}
        \mathrm{C}_{m_1}^{j - (m_2 - m_1)}}{(2K-1)\mathrm{C}_{K}^n}\\\\
\end{split}.
\end{align} 
\end{proposition}

\begin{IEEEproof}
See Appendix \ref{app:A}.
\end{IEEEproof}

\color{black}
We note here that with all other parameters fixed, $\Pplr$ as a function of $n$ has a minimum. 
However, since $n$ is an integer and the formula in (\ref{eq:PLRraw}) is complicated, we discard discussions on conditions of optimal $n$. In practice, we can simply calculate $\Pplr$ numerically for purposes of system design.
\color{black}

In general, increasing the number of \ac{SC}s that a \ac{UE} accesses can drastically reduce packet loss rate. However, from a geometric perspective, the generic argument is that $\left|\mathcal{B}_u\right|$ is limited by $B/N_\bs$, the number of spatial directions in the main lobe of an analog beam. This can also be seen from (\ref{eq:freq_sym}) that \ac{SC}s outside of the main lobe of $\bstaru$ have large gain loss. 

\subsection{Latency and Rate Calculation}

With both major causes of protocol overhead addressed, in this subsection we discuss the performance metrics used in the evaluation of the system. Since the system is designed to fulfill stringent latency requirements, we mainly focus on the overall latency. Based on latency analysis, we then derive the formulas for effective data rate of \ac{UE}s.

As illustrated by \Cref{fig:lat_concept}, the overall latency $\Tl$ can be computed as follows: 
\begin{equation}\label{eq:lat_calc}
\begin{split}
 & \Tl = \Tsync + \Tcont, \\
 & \Tsync = \Tactivation + \Tfailure + \Tdl,\\
 & \Tcont = \Tpacketloss + \Tdl.\\
\end{split}
\end{equation}
Specifically, $\Tactivation$ accounts for the random offset to the nearest broadcasting when a \ac{UE} activates. $\Tfailure$ accounts for unsuccessful synchronization attempts which might include multiple frames. The extra $\Tdl$ is the overhead for successful synchronization. In $\Tcont$, $\Tpacketloss$ stands for packet loss in contention and the extra $\Tdl$ stands for the overhead receiving \ac{ACK} from \ac{DL} broadcasting.

The equation to calculate the effective rate is:
\begin{equation}\label{eq:effR}
    R_{\mathrm{eff}} = \frac{ N_{\mathrm{payloads}}}{\Tcont}
    = \frac{\ggroup N_{\mathrm{packet}}}{\Tcont}
\end{equation} 
where $N_{\mathrm{payloads}}$ is the total payload  and $N_{\mathrm{packet}}$ is the payload per packet in terms of OFDM symbols.

\color{black} The most non-trivial trade-off in our proposed design is controlled by $\ggroup$.  On one hand, as indicated by the numerator in (\ref{eq:effR}), using a large number of SCs per RB (large $\ggroup$) boosts the data rate of \ac{UE}s. On the other hand, large $\ggroup$ also enlarges the denominator in (\ref{eq:effR}) through (\ref{eq:lat_calc}), (\ref{eq:PLRraw}). The interpretation is that given a fixed $\left|\mathcal{B}_u\right|$, increasing $\ggroup$ enhances useful payloads but also reduces the number of \ac{RB}s which means \ac{UE}s become more vulnerable to collisions.
\color{black}

Since the contention overhead depends on the number of users participating in contention at that time, it is hard to evaluate it analytically. Simulation results will analyze these performance metrics in the next section.

%

\section{Numerical and Simulation Results}
\label{sec:Numer}

In this section, we support our analysis in the previous sections with numerical results. 
 \color{black} We start with a brief illustration on how certain system parameters are chosen in the simulations.
\begin{itemize}
    \item{\textit{Size of \ac{BS} antenna $N_{\mathrm{B}}$}: As discussed in Section~\ref{sec:network_analysis}, $N_{\mathrm{B}}$ limits the 
    size of the set $\mathcal{B}_u$, i.e., \ac{SC} that the $u$-th user uses. In simulation, the number of antenna is set to $N_{\mathrm{B}} = 64$. This enables the \ac{UE}s to access at most $\left|\mathcal{B}_u\right|\leq B/N_{\mathrm{B}} = 32$ \ac{SC}s in its narrow baseband within 3dB gainloss with $B = 2048$ as specified by the number of points in \Cref{tab:numer1}. The \ac{UE}'s  baseband bandwidth is then approximately 16MHz.}
    \item{\textit{Grouping factor $\ggroup$}: 
         To keep the discussion simple, we assume that for each user, the number of resource blocks $K$ in $\mathcal{B}_u$ is an integer. This means $\ggroup$ would be a factor of $\left|\mathcal{B}_u\right|$. In simulations, we consider cases where $\ggroup = 1,2,4$. Using $
    \ggroup = 8$ can greatly enhance the rate of \ac{UE}s but there will be only 4 \ac{RB}s in $\mathcal{B}_u$ to avoid collisions.}
    \item{\textit{User active rates $p$}: 
        For the ease of implementation in our simulations, $p$ exclusively controls how many active users are added per frame. Specifically, in each frame we consider $1000$ new links uniformly distributed in a semi-circle covered by the rainbow beam pattern. Activations of these new \ac{UE}s are treated as independent events, i.e., 
    $U$ as a random variable follows binomial distribution $U \sim B\left(1000, p\right)$. These new users will contend with  persistent users added in a previous frame.
    }
\end{itemize}
 \color{black}

\color{black}
Given $N_{\mathrm{B}}$, specifications of link budget at different distances are given in \Cref{tab:lbnoise1}. The network is assumed to operate at \SI{60}{\giga\hertz} carrier frequency at which free space path loss model in \cite{6824746} is adopted with \ac{LoS} pathloss exponent equal to $2$. Specifically, \ac{DL} noise power is calculated based on \ac{BS}'s active bandwidth of \SI{1}{\giga\hertz}, i.e., a \ac{UE}'s receiving \ac{SNR} is 
independent of its active bandwidth. 

\begin{table} [htbp!]
\caption{\ac{DL} Link Budget  Calculation Assuming 1GHz Effective Noise Bandwidth}
\centering
\begin{tabular}{|c|ccc|}
\hline 
Parameter & \multicolumn{3}{|c|}{Value}\tabularnewline
\hline
Distance [m] & 100 & 200 & 400\tabularnewline
\hline 
 \multicolumn{4}{|c|}{BS Configuration}\tabularnewline
Tx Power [dBm] & \multicolumn{3}{|c|}{20} \tabularnewline
Antenna Gain [dBm] & \multicolumn{3}{|c|}{18.1} \tabularnewline
\multicolumn{4}{|c|}{Channel}\tabularnewline
Path loss [dB]  & 108 & 114 & 120\tabularnewline
\multicolumn{4}{|c|}{Rx Power Configuration}\tabularnewline
Rx power w/o BF [dBm] & -69.9 & -75.9 & -81.9 \tabularnewline
Noise Figure & \multicolumn{3}{|c|}{12} \tabularnewline
Noise Power & \multicolumn{3}{|c|}{-83.9} \tabularnewline
Shadow Fading & \multicolumn{3}{|c|}{4.2} \tabularnewline
Rx DL SNR [dB] &  -2.2 & -8.2 & -14.2 \tabularnewline
\hline 
\end{tabular}
\label{tab:lbnoise1}
\end{table}

With the radius of coverage, the activation rate $p$, and the current frame length (specified in
\Cref{tab:linkbud}),  one can compute user capacity of the system. For instance, a \SI{400}{\meter} radius of coverage and $p = 0.03$ gives the following density on average:
\begin{align}
\begin{split}
    \rho & = \frac{\mathbb{E}\left[U\right]}{\pi \times 400^2 \times 0.5}\times \frac{1}{T_{\mathrm{frame}}} \\
    & = \frac{1000 \times 0.03}{\pi \times 400^2 \times 0.5}\times \frac{1}{125 \times 10^{-6}}\\
    & \approx 0.95 \\
\end{split}
\end{align}
in unit of \ac{UE} activation per \SI{}{\meter}$^2$ per second. This density $\rho$ serves as a generic metric that evaluates network capacity.

A perceivable drawback of the proposed system is the low spectral efficiency. The drawback comes from 
the following facts:
\begin{itemize}
    \item UE packets are short such that good coding schemes are not applicable. Non-payload portions cannot be ignored. 
    \item \ac{UE}s trade its bandwidth for low rate of collisions.
\end{itemize}

Even if $\ggroup = 4$ is employed along with $16$-\ac{QAM} and $1/2$ coding scheme, maximum $R_{\mathrm{eff}}$\footnote{the maximum rate is when a \ac{UE} always succeeds in its first attempt of contention} for a single \ac{UE} is merely $\frac{4\times 18}{125\times 10^{-6}}\times 4\times \frac{2}{3}= 1.536$ Mbps. Thus the proposed systme is not intended for high rate applications.
 \color{black}


In the evaluation we use the following parameters unless otherwise specified. The \ac{BS} has an array size of $N_{\bs}= 64 $ elements. The total number of \ac{SC} is $B = 2048$ and \ac{UE}s operate on $\left|\mathcal{B}_u\right| = B/N_\bs = 32$ \ac{SC}s. 
A maximum \ac{UE} transmit power is set to \SI{23}{\dBm} which provides a coverage of $400$m to achieve about \SI{10}{\decibel} \ac{SNR} for \ac{UL} transmission. For users near the cell edge, due to their very low \ac{SNR}, multiple synchronization attempts might be needed to ensure a satisfying detection performance. We  evaluate the probability of successful synchronization as a function of \ac{SNR} and its impact on access latency. 

\begin{figure}[htbp!]
    \centering
    \includegraphics[width=0.45\textwidth]{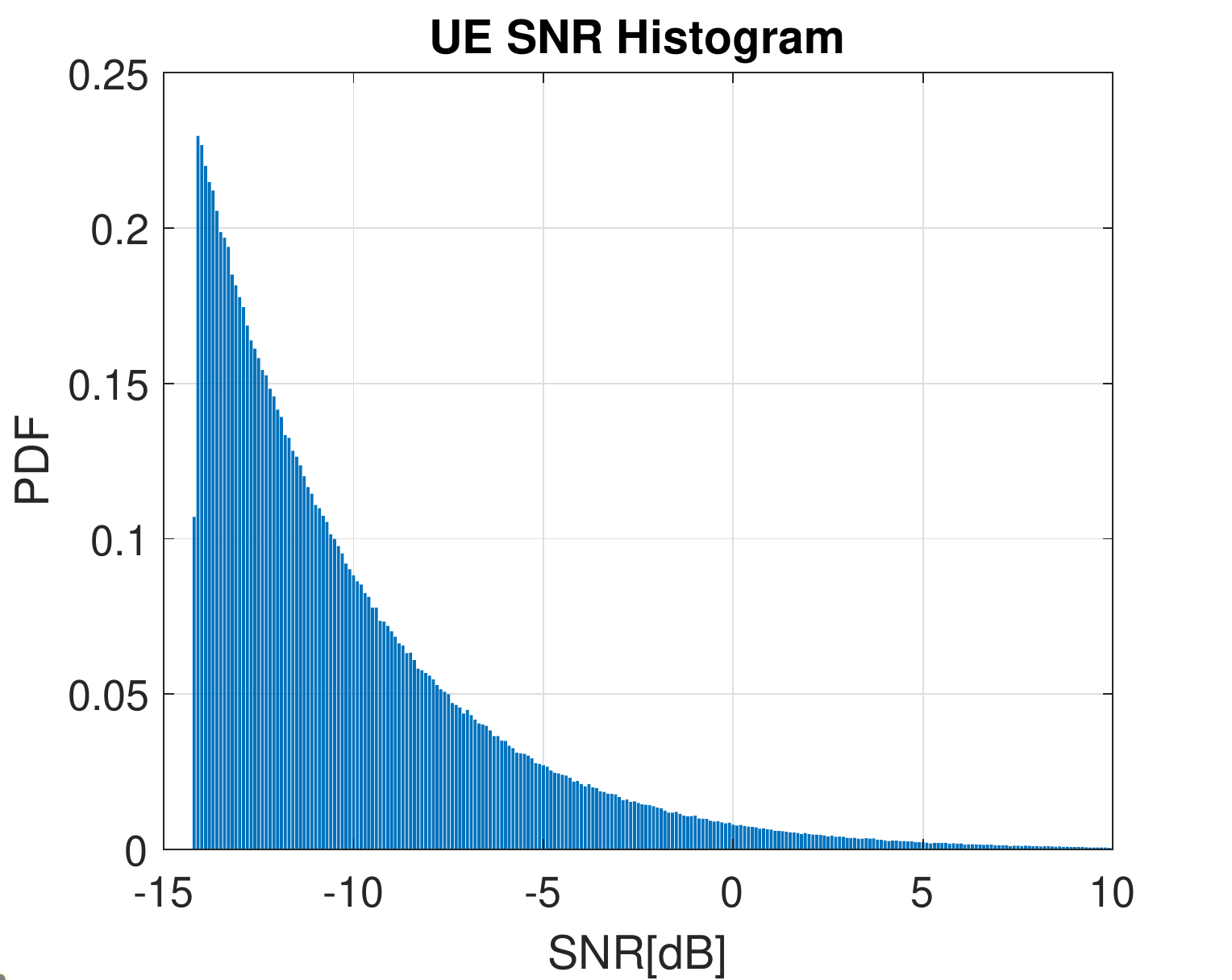}
    \vspace{-2mm}
    \caption{2D uniform user distribution in a semi-circle with $400$m radius. }
    \vspace{-2mm}
    \label{fig:snr_hist}
    \end{figure}

An overall histogram of \ac{DL} \ac{SNR} values for UEs randomly distributed in a semi-circle of \SI{400}{\meter} radius is given in \Cref{fig:snr_hist}. This range of \ac{SNR} values are used throughout simulations.

\subsection{Synchronization Performance}
\begin{figure}[htbp!]
    \centering
    \includegraphics[width=0.45\textwidth]{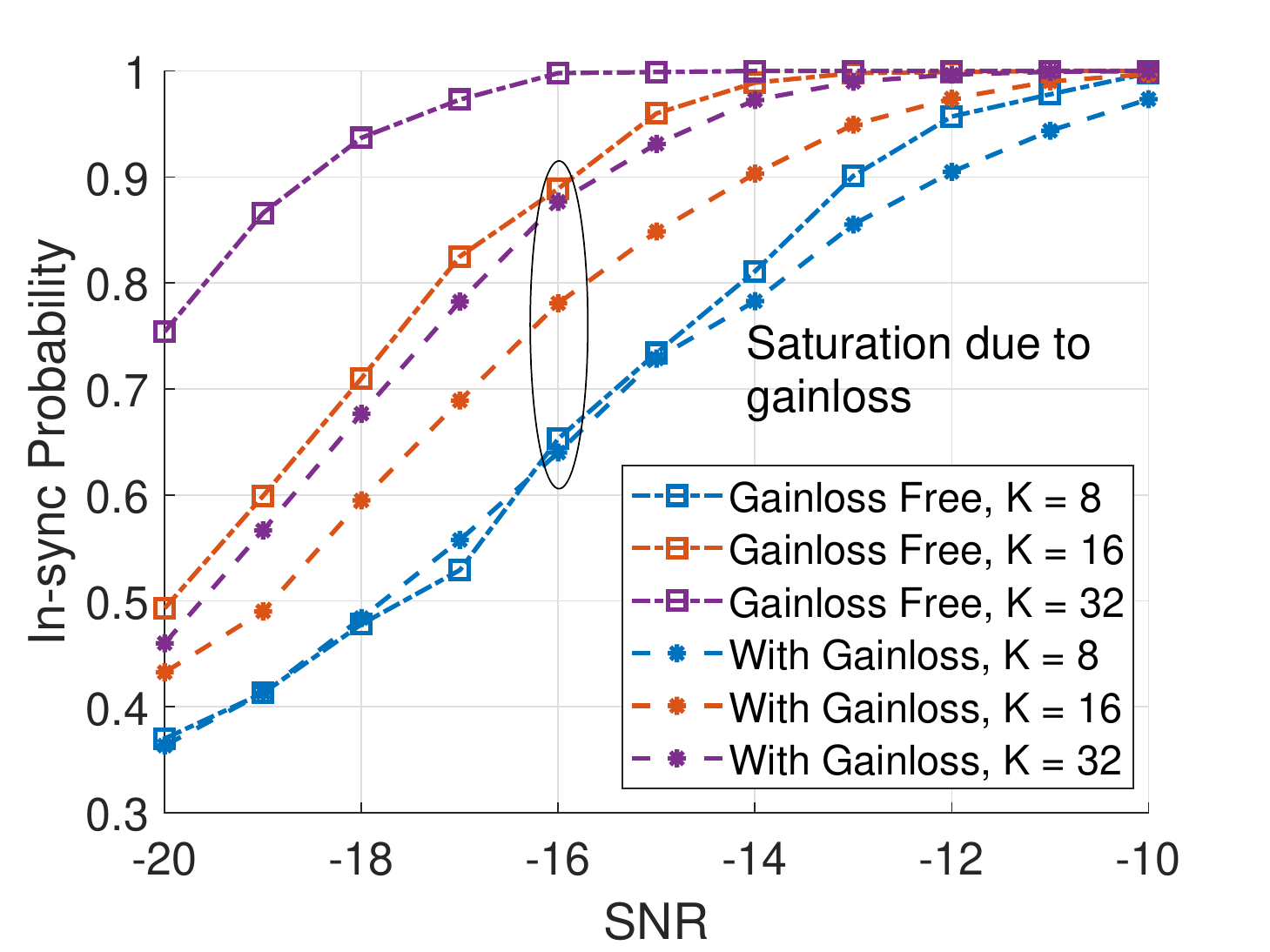}
    \vspace{1mm}
    \caption{\ac{DL} detection with $L = 22$ symbols, with $B = 2048$ \ac{SC}s and $N_\bs = 64$ antenna elements, $B/N_\bs$ = 32.}
    \vspace{-2mm}
    \label{fig:sync_performance}
    \end{figure}

    Through Monte Carlo simulations, we explore probability  of correct estimation of $D_u^*$ as a function of broadband SNR.
    \Cref{fig:sync_performance} gives detailed comparison of synchronization performance with and without including impacts of the aforementioned gain loss and a multiple-path channel. As specified in \Cref{tab:linkbud},
    a \ac{PN} sequence with length $L = 22$ is employed as preamble. 
    
The results in \Cref{fig:sync_performance} clearly show that the gain loss has a non-negligible impact. The more \ac{SC}s are used for narrowband transmission, the larger the impact of beamforming gain loss is. \color{black} 
As $\left|\mathcal{B}_u\right|$ approaches $B/N_\bs$, the synchronization performance starts to saturate.
\color{black}

With only $\left|\mathcal{B}_u\right| = 8$ \ac{SC}s in band, probability of successful synchronization is not visibly impacted by the rainbow beam pattern. This means that with $ \left|\mathcal{B}_u\right| \leq B/4N_\bs$ , \ie, only a small number of \ac{SC}s included in $ \mathcal{B}_u$, the impact of rainbow beamforming is trivial. 
With 16 or 32 \ac{SC}s in band, synchronization performance loss due to distortion becomes noticeable. There is a \SI{4}{\decibel} equivalent SNR loss due to distortion when system uses  $\left|\mathcal{B}_u\right| = 32$ \ac{SC}s, which is effectively the threshold for $\left|\mathcal{B}_u\right|$ as discussed previously. In such case, the detection performance is merely comparable with the case of $\left|\mathcal{B}_u\right| = 16$ when there is no distortion.
%
%

The implications of numerical results are the following. Firstly, the proposed narrowband synchronization beacon is robust to both thermal noise and frequency selectivity introduced by the rainbow beam. In a typical use case of \ac{mmW} \ac{IoT} with large number of devices, high probability of \ac{DL} synchronization can be achieved. Secondly, synchronization performance would benefit greatly from an accurate estimation of $\bstaru$ where (\ref{eq:syncUE}) can then be altered to count for the gain loss coherently. Otherwise, our analysis suggests that $\left|\mathcal{B}_u\right| \leq B/4N_\bs$ can be a generic threshold for achieving negligible beamforming gainloss.

\subsection{Collision and Packet Loss Rate}

\begin{figure}[htbp!]
  \centering
  \begin{tabular}[b]{c}
    \includegraphics[width=0.95\linewidth]{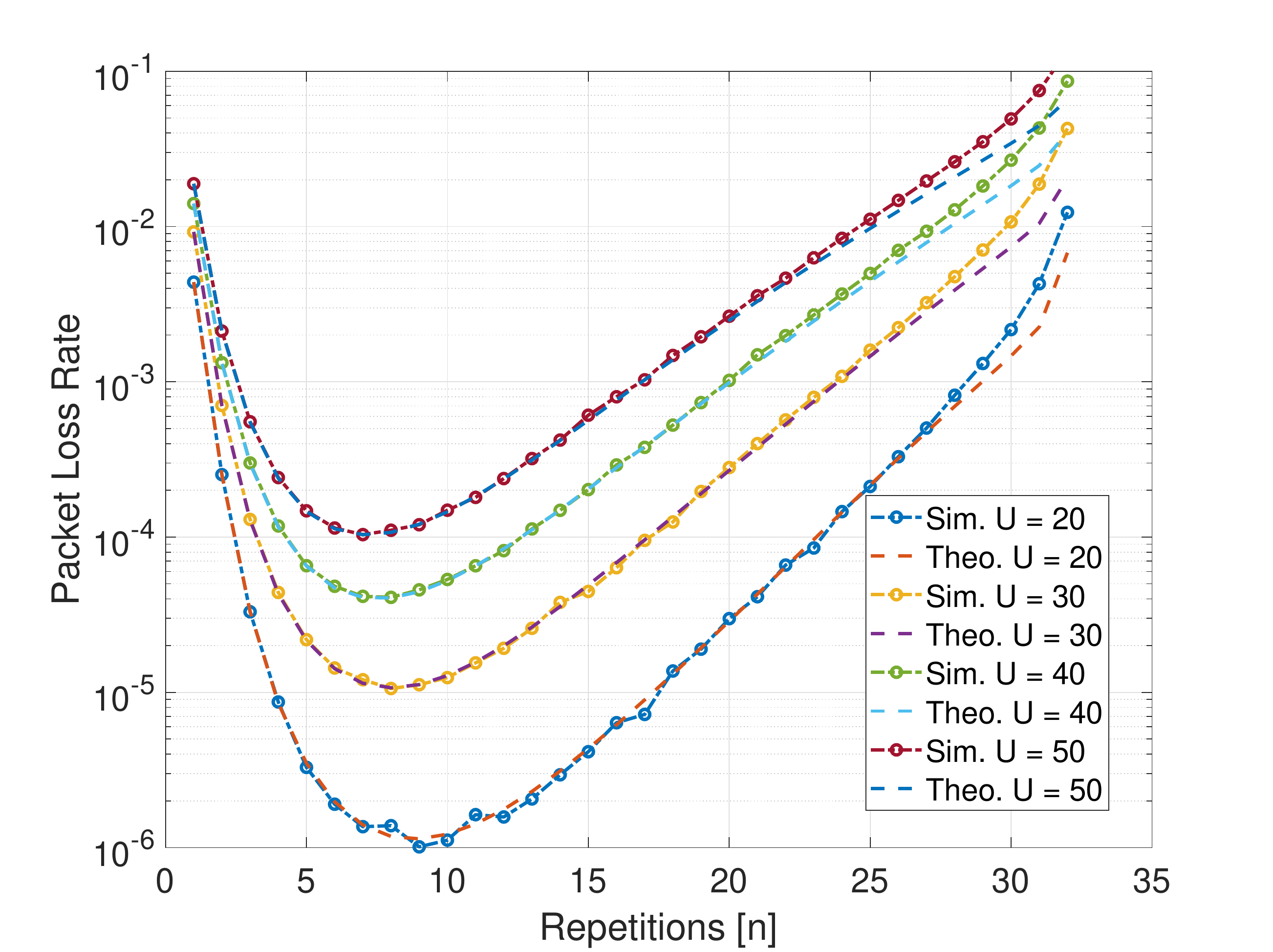} \\
    \small (a) Packet loss rate with UE's bandwidth as wide as $K = 32$ \ac{RB}s.\\
    \includegraphics[width=0.95\linewidth]{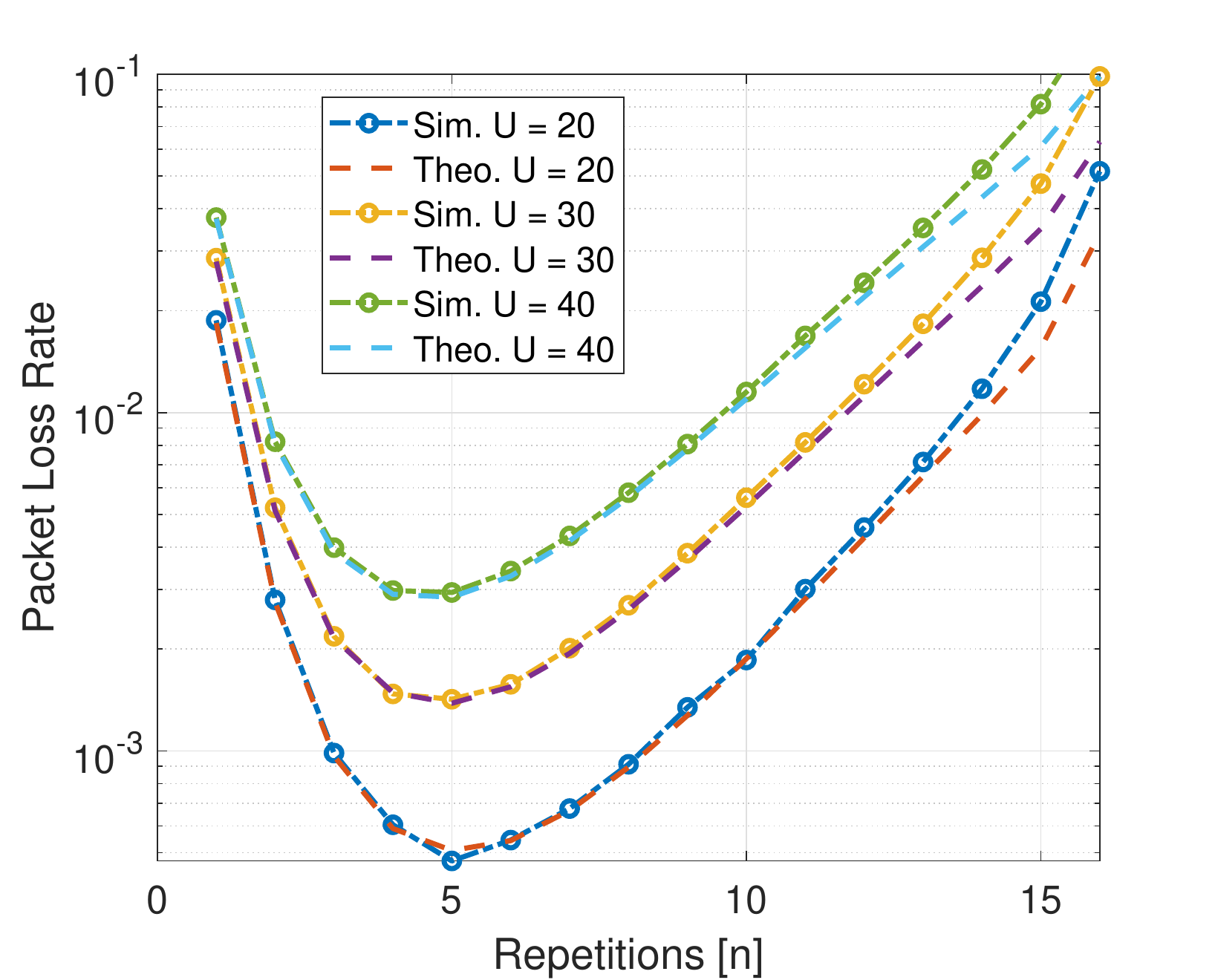} \\
    \small (b) Packet loss rate with UE's bandwidth as wide as $ K = 16$ \ac{RB}s.
  \end{tabular}
      \caption{Packet loss rate $\Pplr$ as function of UE's repetition transmission number $n$. Different curves are evaluated with different number of active users $U$ in contention. \color{black} The optimal $n$ drifts slowly to smaller values as $U$ increases.\color{black}}
  \vspace{-1mm}
\label{fig:plr_performance}
\end{figure}

 \Cref{fig:plr_performance} presents both the simulated and theoretical packet loss rate given by (\ref{eq:PLRraw}). 
 In the evaluation, we focus on the two cases $\ggroup = 1,2$.

We have the following observation from the results. Firstly, the theoretical results agree well with simulations for small and moderate $n$ values. This confirms the correctness of the Proposition 1 where we derived \ac{PLR}. Admittedly, discrepancy occurs for large $n$ values because our derivation omitted the cases when each user uses repetition transmission that occupies the entire band. Secondly, the results indicate that for a given number of active users, there is an optimal value of repetitions that minimizes $\Pplr$. The optimal value of $n$ decreases with an increasing number of active users in contention. This makes sense since intuitively users should avoid making too many repetitions when the network is crowded, so as to leave resources for others. Lastly, compared to probability of failed synchronizations, packet losses due to collisions among users have much smaller probabilities in the given network settings. Results from \Cref{fig:plr_performance} show that even with $\ggroup = 2$ and with 20 active users transmitting simultaneously, the packet loss rate is still below $10^{-3}$. With multiple retransmissions, the system can then provide high reliability.

\subsection{Latency and Rates}

Next we evaluate the total latency $\Tl$ due to \ac{DL} synchronization and \ac{UL} contention. \color{black} Here we focus on the trade-off based on $\ggroup$ as specified in IV-C. \color{black}  We note here again that users are generated with uniform distribution in a 2D-sector of \SI{400}{\meter} radius and activation probability of $0.03$ per frame over $1000$.

 \begin{figure}[htbp!]
  \centering
  \begin{tabular}[b]{c}
    \includegraphics[width=0.95\linewidth]{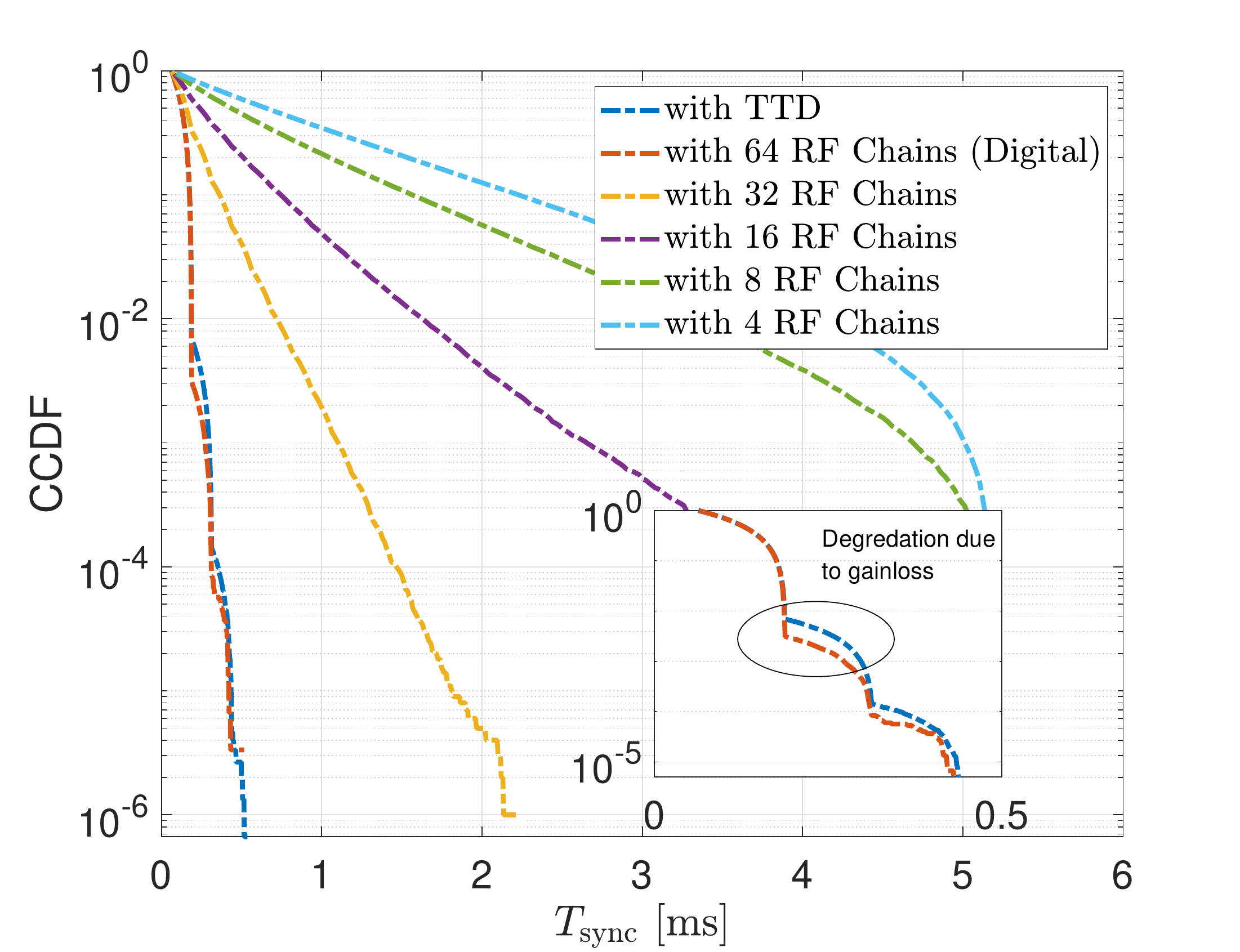} \\
    \small (a) Statistics for $\Tsync$ compared with phased array\\
    \includegraphics[width=0.95\linewidth]{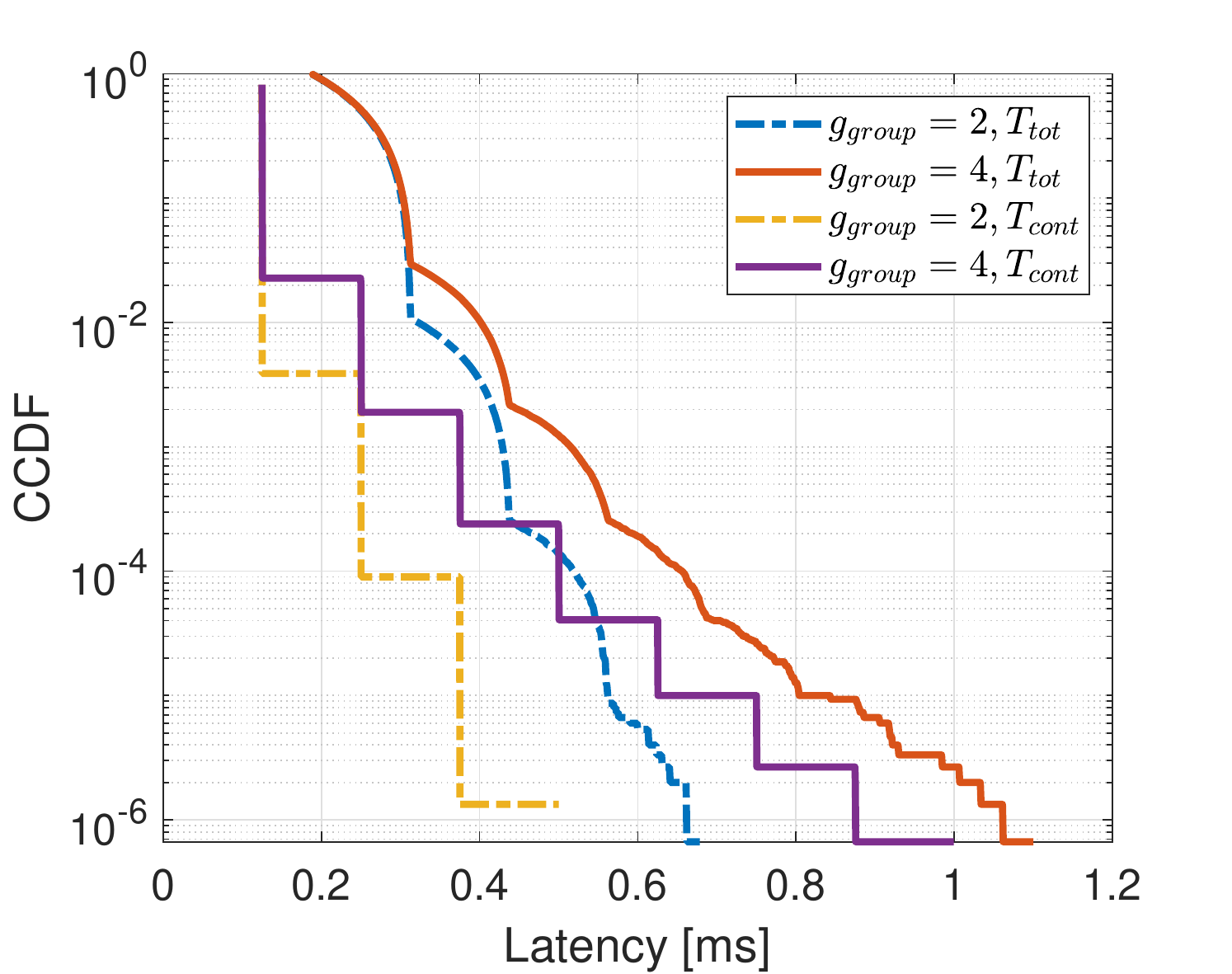} \\
    \small (b) Statics for $\Tl$ with different sizes of \ac{RB}s.
  \end{tabular}
      \caption{Simulated statistics of latency with active rate $p = 0.03$. $\Tcont$ exhibits a typical zig-zag pattern due to multiple contention attempts.}
  \vspace{-1mm}
\label{fig:lat_performance}
\end{figure}

\color{black}
Simulation results for user-plane latency are presented in \Cref{fig:lat_performance}. In \Cref{fig:lat_performance}a, a comparison on synchronization overhead between the proposed \ac{TTD} array and phased arrays with multiple RF chains is presented. In both cases, \ac{UE}s operate on only 32 \ac{SC}s and \ac{BS} has a 64-element array. To make a fair comparison, in the case of phased array \ac{BS} employs random beamforming strategy. In each frame, it steers analog beams randomly to serve users. Thus in both cases, there is no overhead due to control signaling or scheduling\footnote{Presumably, the number of beams that a BS can steer with the phased array is limited by the number of RF chains. Since all RF chains share the same band resources, BS should point them to distinct angular sectors to cover as many users as possible. In our comparison all sectors have a equal probability to be chosen. To our best knowledge, this is the only regime with phased array that's compatible with a grant-free multiple access scheme as depicted in \cref{fig:lat_concept}}. From the result, the performance of the proposed system approaches that of a fully-digital array. As discussed in the previous section, the degradation comes from beamforming gainloss due to frequency dependent beam pattern. For  conventional phased arrays, as the number of \ac{RF} chains decrease, the overall beam coverage is limited as depicted in \Cref{fig:my_label}. \ac{UE}s start to experience prohibitively long synchronization overhead that cannot satisfy requirements of latency-critical uses cases. 
\color{black}

The \Cref{fig:lat_performance}b shows that with $2$-grouping and active rate $p = 0.03$, $\Tsync$ dominates the overall latency while with $4$-grouping, $\Tcont$ dominates  the overall latency. The tail in contention latency is due to persistent, spatially clustered users in certain narrow band segment. Although in both cases the general \ac{URLLC} requirement for reliability greater than $1 - 10^{-5}$ at \SI{1}{\milli\second} latency is satisfied, in $4$-grouping case the margin is small.
This means that $2$-grouping can still support more active users per frame (higher active probability $p$) while in $4$-grouping case $p = 0.03$ is nearly the maximum activity rate it can support 
for \ac{URLLC}. 

\begin{figure}[htbp!]
  \centering
    \includegraphics[width=0.95\linewidth]{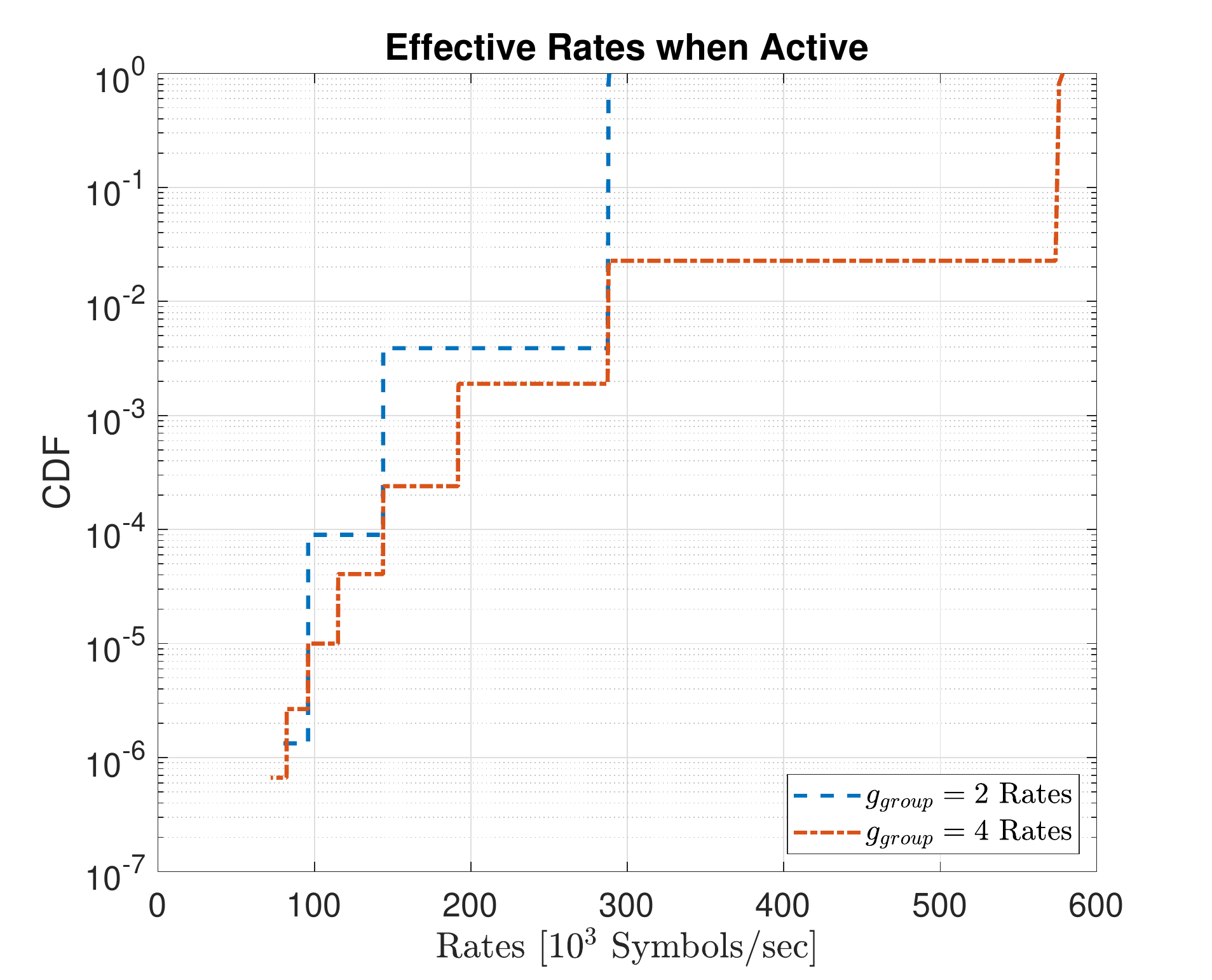}
  \caption{Rate of a single \ac{UE} with active probability $p= 0.03$ based on contention delay $\Tcont$ shown in \Cref{fig:lat_performance}. Actual rates in bits per second depend on modulation and coding schemes. }
\label{fig:rate_performance}
\end{figure}

In \Cref{fig:rate_performance}, we show the effective rates for different grouping strategies. For instance, subcarrier grouping with $\ggroup = 4$ achieves higher rates than $\ggroup = 2$ yet users are more likely to collide. As stated before, the proposed design is tailored for \ac{URLLC} and for sporadic, latency critical transmissions. Higher grouping ratio $\ggroup$ might bring higher rates, but is very unlikely to satisfy \ac{URLLC} requirement.


\begin{figure}[htbp!]
  \centering
  \begin{tabular}[b]{c}
    \includegraphics[width=0.95\linewidth]{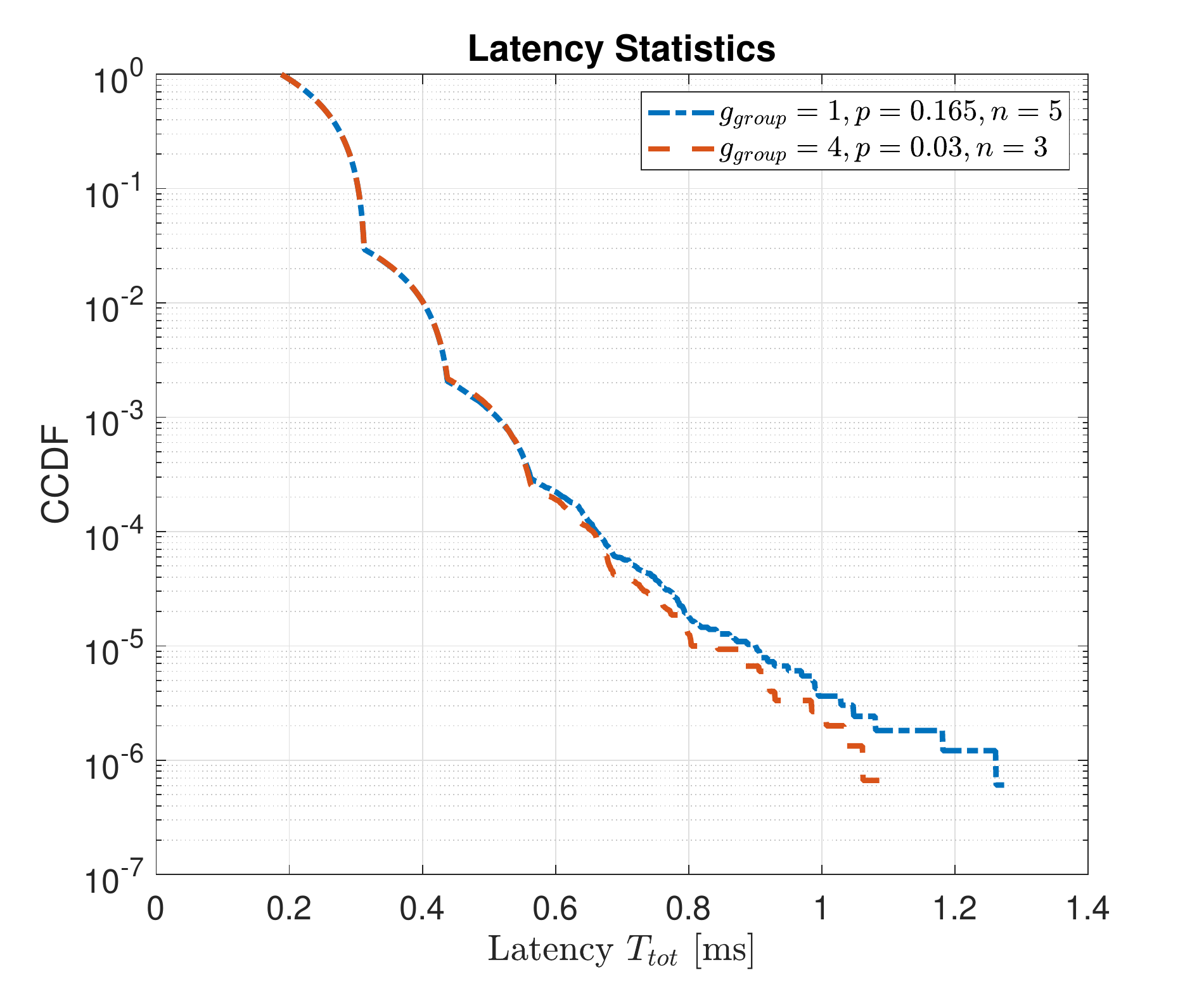} \\
    \small (a) Nearly Saturated Latency Statistics\\
    \includegraphics[width=0.95\linewidth]{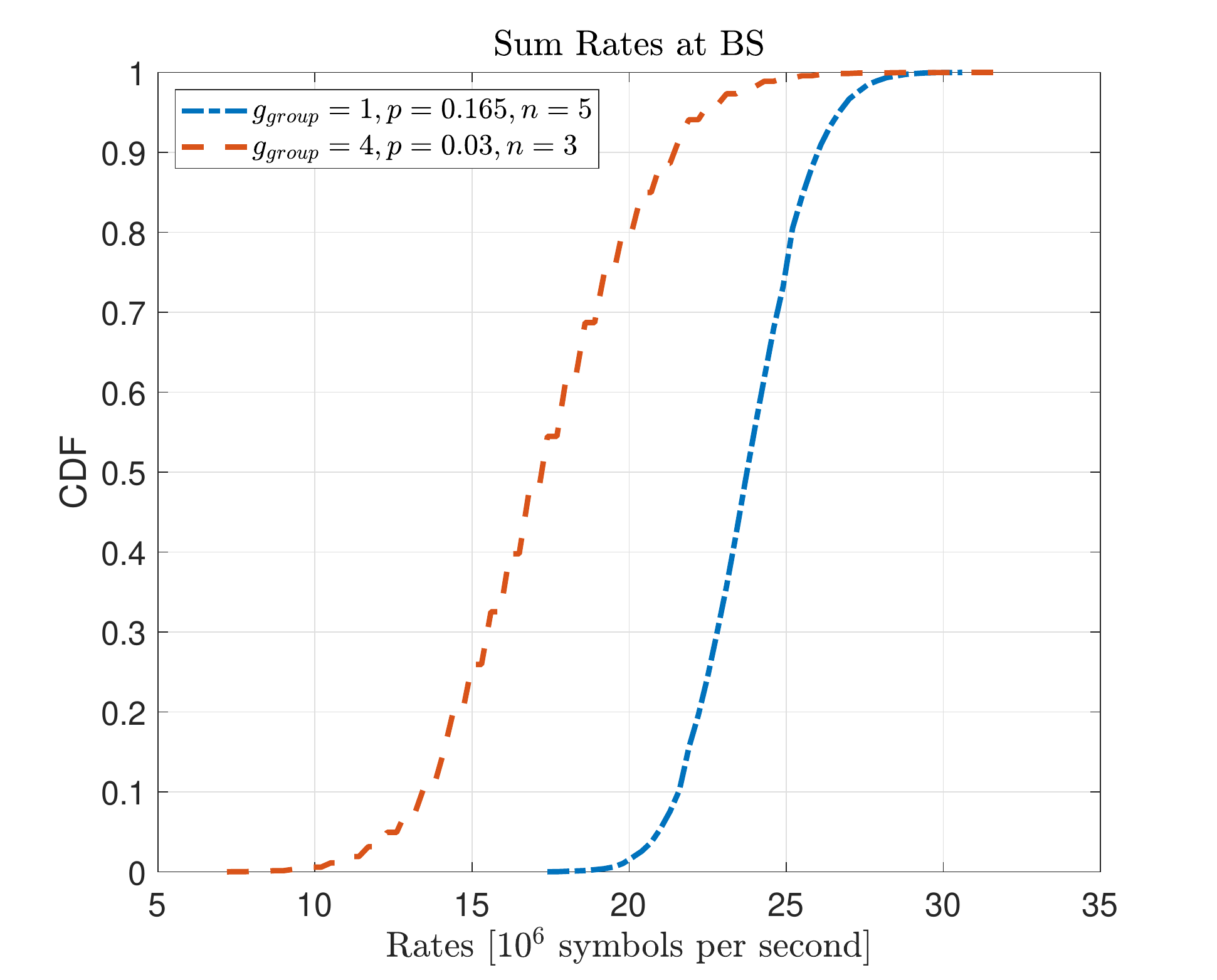} \\
    \small (b) Aggregated Throughput
  \end{tabular}
  \caption{Latency and rates with different active probabilities. The $p = 0.03$ curve is the same as that in previous results}
\label{fig:sumrate_performance}
\end{figure}

\color{black} In \Cref{fig:sumrate_performance}, we demonstrate trade-off of $\ggroup$ with respect to network capacity. Specifically, we gradually increase $p$ to see what's the approximate maximum traffic that the network can support with $\ggroup = 1$ and $\ggroup = 4$, respectively. The saturation for $\ggroup = 1$ appears at about $p = 0.165$ and as mentioned earlier the saturation for $\ggroup = 4$ appears at about $p = 0.03$. Using \Cref{eq:PLRraw}, we numerically calculate $\Pplr$ and find that the optimal number of repetitions $n$ are $5$ and $3$, respectively. Although with $\ggroup = 1$, each user has only one forth of the data rate, \ac{BS} can serve 5 times as many users as that in the case of $\ggroup = 4$. \color{black}

In both cases, the achievable latency for a reliability of $1 - 10^5$ is very close to \SI{1}{\milli\second}, as can be seen from \Cref{fig:sumrate_performance}. Although in the first case there is no grouping and rate of each user is small, the system can still achieve a higher sum rate on average by serving more users. With $p = 0.165$, a $16$-\ac{QAM}, and $1/4$ coding rate, the aggregated throughput can reach \SI{28}{\mega\bps} Mbps on \ac{BS} side. In this case, the BS serves on average $165$ active, low-rate users per frame. Thus the proposed system is better utilized when handling massive multiple access with stringent latency and reliability requirements.

\section{Discussion and Future Work}
\label{sec:discussion}
In this section, we briefly discuss the implications of numerical results and open research questions.

The narrow bandwidth and number of SCs used by the UE are critical parameters for system performance. With the increased number of SCs that UE can access, there is improvement in both synchronizations and contentions. On the other hand, the physical nature of rainbow beamforming limits the number of \ac{SC}s (approximately $B/N_\bs$) that a user can access: only a fraction of radio resources are aligned with the \ac{AoA} of a \ac{UE} and \ac{SC}s at edge of the segment $\mathcal{B}_u$ suffer severely from beamforming mismatch. 
 
Based on the presented analysis, the proposed system has the following performance: 
\begin{itemize}
    \item{\textit{Power and Coverage}: The proposed rainbow link based network can achieve a coverage up to \SI{400}{\meter} when \ac{BS} with $64$ \ac{TTD} analog antenna array serves single antenna \ac{UE}s with the transmit power of \SI{23}{\dBm} assuming line-of-sight condition.} 
    \item{\textit{Rates and user capacity}: The BS can serve up to 5 active UEs per second per \SI{}{m}$^2$ with Mbps data rates depending on the grouping of SCs per RB.} 
    \item{\textit{Latency and reliability}: The UEs experience \SI{1}{\milli\second} latency and reliability higher than $1 - 10^{-5}$  when the optimal number of repetitions $n$ is used.}
\end{itemize}

The greatest strength of the proposed system is its flexibility to support combined \ac{URLLC} and massive multiple access. Based on the proposed design of numerology and multiple access protocol, the system is better suited to serve a massive number of users than to provide high data rates. When conservative grouping is used (in our case no grouping or 2-grouping), the proposed system can eventually achieve a sum rate at \ac{BS} higher than that of the case with larger grouping. 

The open research questions include but are not limited to the following:


\textit{
    Efficient Frame Design:
} If no grouping is used, one frame has only $18$ symbols in its payload based on the proposed frame design. Typical \ac{URLLC} packets have 32 bytes, so either dense modulation scheme  or more aggressive grouping strategies would have to be used. In this work we assume $1:1$ \ac{UL} \ac{DL} ratio. Alternative frame design, for instance, a frame length of \SI{250}{\micro\second} with $1:3$ \ac{UL} and \ac{DL} ratio could support higher data rate for each user with extra \ac{DL} \ac{TTI}s all contributing to the payload. 



\textit{\color{black} Robustness for Non-\ac{LoS} channels and user mobility}: In the current discussion, \ac{UE}s are assumed to be in \ac{LoS} channels. But the same discussion might not be applicable for non-\ac{LoS} channel conditions. For instance, when a \ac{UE} suffers from blockage, it loses track of its anchor \ac{SC} and needs to switch to a different path. As pointed out in III-A, the switching of $\mathcal{B}_u$ without any prior requires prohibitively long overheads. Thus the current design is not robust to mobility of \ac{UE}s and non-\ac{LoS} channel conditions. These questions need to be addressed for practical implementations of rainbow-link. \color{black} 

\textit{Dynamic Control of Random Access:} Increasing bandwidth of the narrowband \ac{UE}, at the expense of the hardware cost,  can greatly boost performance of the proposed 
system. While we analyzed the system performance under uniform user distribution, distributions of users based on clustered model could further intensify contentions. In this case there might be no optimal number of  repetitions as the UEs would persistently contend with other UEs in the same spatial sector.
To overcome this effect, the BS can control the number of repetitions that each user can make. If there are too many collisions, \ac{BS} would tell each UE to reduce its repetitions so as to make sure that at least some UEs can transmit.

\textit{Alternative Multiple Access Protocol:} The proposed system can benefit from non deterministic repetition coding strategy. As it has been proposed in  \cite{5668922}, each user can use a random number of repetitions $n$ for each contention period. This might increase system robustness to clustered scenario. Unfortunately, the analysis used in \cite{5668922} is
not directly applicable due to its assumption of sufficient length of code words. Alternatively, instead of repetition coding, a non-orthogonal coding can be applied. Each user can transmit data on all \ac{SC}s in its band and multiply data symbols with a unique code in frequency domain. Though collisions will happen on \ac{SC}s, \ac{BS} can still decode with these unique code words. This is similar to \cite{7862785} where NOMA is employed for multiple access.

%
%
\section{Conclusion}
\label{sec:Conclusion}

In this work we proposed a novel frequency domain multiple access, referred as rainbow link, that exploits wideband spectrum at \ac{mmW} to serve a large number of narrowband users. By exploiting TTD array architecture at the BS,  frequency resources are mapped to specific spatial directions so that users can be assigned a subset of SCs in \ac{OFDMA} and leverage beamforming gain of the entire array. Rainbow link enables a grant free multiple access and can support a very large number of users with stringent latency requirements. With a single \ac{RF} chain and 64-element array, the \ac{BS} can provide reliable access with \SI{400}{m}
\ac{LoS} coverage and sub-millisecond overhead for up to 5 \ac{UE} activations per second per square meter. We believe the proposed rainbow-link can be a candidate for future critical \ac{mMTC} use cases.

\bibliographystyle{IEEEtran}
\bibliography{IEEEabrv,references}

\appendix

\subsection{Generic Formula for Set Covering Distribution}\label{app:A}

\color{black} For the simplicity of notation, in the appendix we use $B$ for the total number of \ac{RB}s so that it integrates $\ggroup$. \color{black}
Before the detailed of derivation of (\ref{eq:PLRraw}), we first rewrite it in a more intuitive manner:

\begin{equation}
\label{eq:PLR_raw}
\begin{split}
    \Pplr(U,K,B,n) &  = \sum\limits_{i = 1}^{U - 1} f\left(U - 1, \frac{2K-1}{B},i\right)\\
    & \times P_{K,n}(i), \\
\end{split}
\end{equation}

$P_{K,n}(i)$ is probability of packet loss of a user given $i$ other users are transmitting on nearby \ac{SC}s in its narrow baseband $\mathcal{B}_u$.  Specifically, nearby \ac{SC}s actually refer to the $\left(2K-1\right)$ \ac{SC}s whose narrow band segments have intersection with 
$\mathcal{B}_u$. Since the user's choice of $n$ \ac{SC}s are completely random, $P_{K,n}(i)$
can be decomposed in a straight forward manner: 
\begin{align}
\begin{split}
    P_{K,n}(i) & = \sum\limits_{ j = 0}^{K} P(j\ \mathrm{among}\ K\ \mathrm{
    \ac{SC}s}\ \mathrm{\ occupied\ by\ }i\ \mathrm{users} ) \\
    & \times \mathrm{C}_{j}^n\left/\mathrm{C}_{K}^n\right. \\
\end{split}
\end{align}
Essentially, packet loss results in the case when all $n$ choices of the user fall on occupied \ac{SC}s. Therefore, once we know the probability distribution of the number of occupied \ac{SC}s, packet loss rate can be thoroughly calculated. The strategy here is to characterize occupancy of the selected narrow band as a Markov chain where each competing user causes transition of states in terms of number of occupied \ac{SC}s.  Let $ \mathbf{p}_i \in  \mathbb{R}^{K+1}$ be the vector characterizing state of the narrow band with $i$ competing users, \ie, $\left[\mathbf{p}_i\right]_j = P\left(j\ \mathrm{among}\ K\ 
    \ac{SC}s\ \mathrm{\ occupied\ by\ } i\ \mathrm{users}\right) $. Correspondingly, let $\mathbf{T} \in \mathbb{R}^{(K+1) \times (K+1)}$ be the transition matrix created by a single competing users. Specifically,
    \begin{align}
    \begin{split}
    \left[\mathbf{T}\right]_{m_1,m_2} & = P\left(\mathrm{new\ m_2 - m_1\ occupied\ \ac{SC}s}\right) \\
    & =\sum\limits_{j = 0}^n\sum\limits_{k = 1}^{K}P\left(\mathrm{user\ occupies\ 
    }\mathrm{band\ with}\ k\ \mathrm{overlap}\right)\\
    & \times P_k\left(\mathrm{user\ adds}\ j\ \mathrm{repetitions\ in\ band}\right)\\
    & \times P\left(j\ \mathrm{repetitions\ add}\ m_2 - m_1\ \mathrm{new}\ \mathrm{occupations}\right) \\
    \end{split}
    \end{align}

Since any segment of joint $K$ \ac{SC}s can be chosen as a narrow band, competing users can have different numbers of overlapping \ac{SC}s with 
the user of interest. As a consequence, for a competing user, its $n$ repetitions might fall only partially on those overlapped \ac{SC}s. Then, those repetitions (on overlapped \ac{SC}s) contributed by a competing user may or may not land on occupied \ac{SC}s in $\mathcal{B}_u$. All these probabilities are given by:

\begin{itemize}
    \item{ Overlapping configurations:
    \begin{displaymath}
    \begin{split}
        P\left(\mathrm{user\ occupies\ 
    }\mathrm{band\ with}\ k\ \mathrm{overlap}\right) & = \\
    \left\{\begin{matrix} 
        2/(2K-1), & 1 \leq k \leq K - 1 \\
        1/(2K-1), &  k = K \\
    \end{matrix} \right.& \\
    \end{split}
    \end{displaymath}
    }
    \item{Probability that a user with a band that has $K$ overlapped blocks with $\mathcal{B}_u$ contributes to $j$ repetitions:
    \begin{displaymath}
        P_k\left(\mathrm{user\ adds}\ j\ \mathrm{repetitions\ in\ band}\right) =
        \frac{\mathrm{C}_{K - k}^{n - j}
        \mathrm{C}_{k}^j}{\mathrm{C}_{K}^n}
    \end{displaymath}
    }
    \item{Probability for newly added repetitions to cause transition: 
    \begin{displaymath}
    \begin{split}
     & P\left(j\ \mathrm{repetitions\ add}\ m_2 - m_1\ \mathrm{new}\ \mathrm{occupations}\right) = \\
     & \qquad  \frac{\mathrm{C}_{K - m_1}^{m_2 - m_1}
        \mathrm{C}_{m_1}^{j - (m_2 - m_1)}}{\mathrm{C}_{K}^j}\\
    \end{split} 
    \end{displaymath}
    This formula is essentially a generic approximation. To our best knowledge, the exact formula here can not be tracked analytically. The main reason is that probabilities for different configurations of $j$ repetitions in $\mathcal{B}_u$ are not even. Competing users with a highly overlapped band might generate some configurations that other users cannot. This means that 
    state of the band can not be fully characterized by the number of occupied \ac{SC}s alone. 
    
    Since a complete treatment is computationally infeasible, we here assume that all configurations for $j$ repetitions are equivalent. This assumption works well for small and medium value of $n$ and introduces deviation for $n$ approaching $K$.}
\end{itemize}
With the transition matrix, calculation of $\mathbf{p}_i$ is relatively simple:
\begin{align}
    \mathbf{p}_i = \mathbf{T}^i\mathbf{p}_0
\end{align} 
And then: 
 \begin{align}
    P_{K,n}(i) = \sum\limits_{ j = 0}^{K} \left[\mathbf{T}^i\mathbf{p}_0\right]_j \mathrm{C}_{j}^n\left/\mathrm{C}_{K}^n\right. 
\end{align}

\begin{IEEEbiography}[{\includegraphics[width=1.0in,height=1.25in,clip,keepaspectratio]{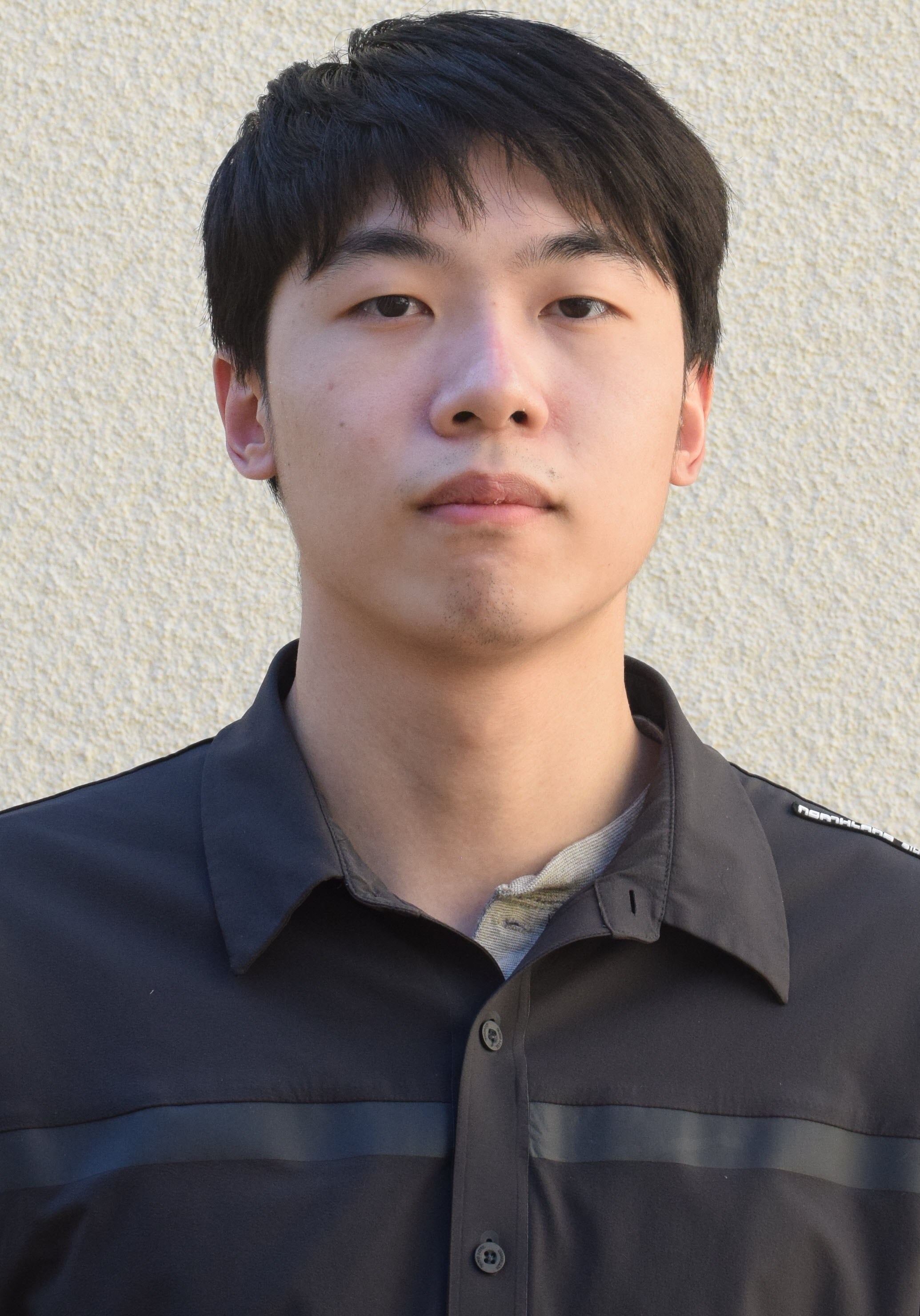}}]{Ruifu Li} received the B.Sc. degree from the University of Wisconsin-Madison in 2020. He is currently working towards his M.S. degree in electrical and computer engineering at University of California, Los Angeles. His research interests include signal processing and wireless communications. 
\end{IEEEbiography}
\begin{IEEEbiography}[{\includegraphics[width=1in,height=1.25in,clip,keepaspectratio]{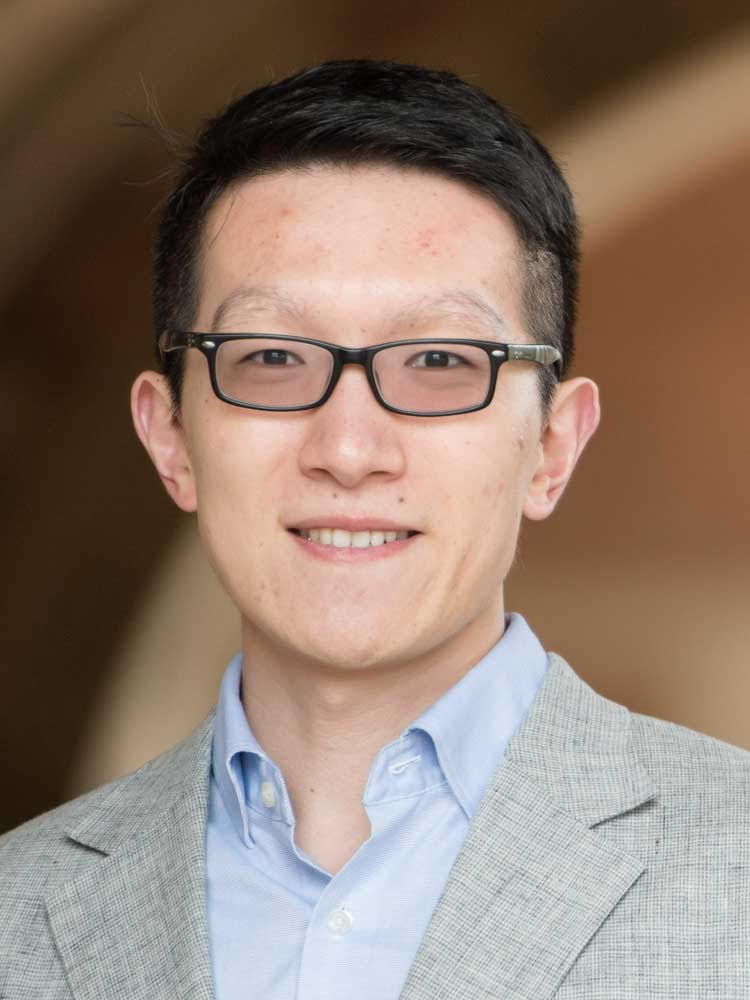}}]{Han Yan}
received the B.E. degree from Zhejiang University, Hangzhou, China, in 2013, and the M.S. and Ph.D. degrees in electrical and computer engineering from UCLA in 2015 and 2020, respectively. He has broad research interests in signal processing and communication system design for millimeter-wave mobile networks, cooperative unmanned aerial vehicles networks, and dynamic spectrum sharing radios. Dr. Yan was a recipient of the UCLA Dissertation Year Fellowship in 2018, Qualcomm Innovation Fellowship in 2019, UCLA ECE Distinguished Ph.D Dissertation Award in 2020, and best paper award at the 2020 ACM mmNets workshop.
\end{IEEEbiography}
\begin{IEEEbiography}[{\includegraphics[width=1in,height=1.25in,clip,keepaspectratio]{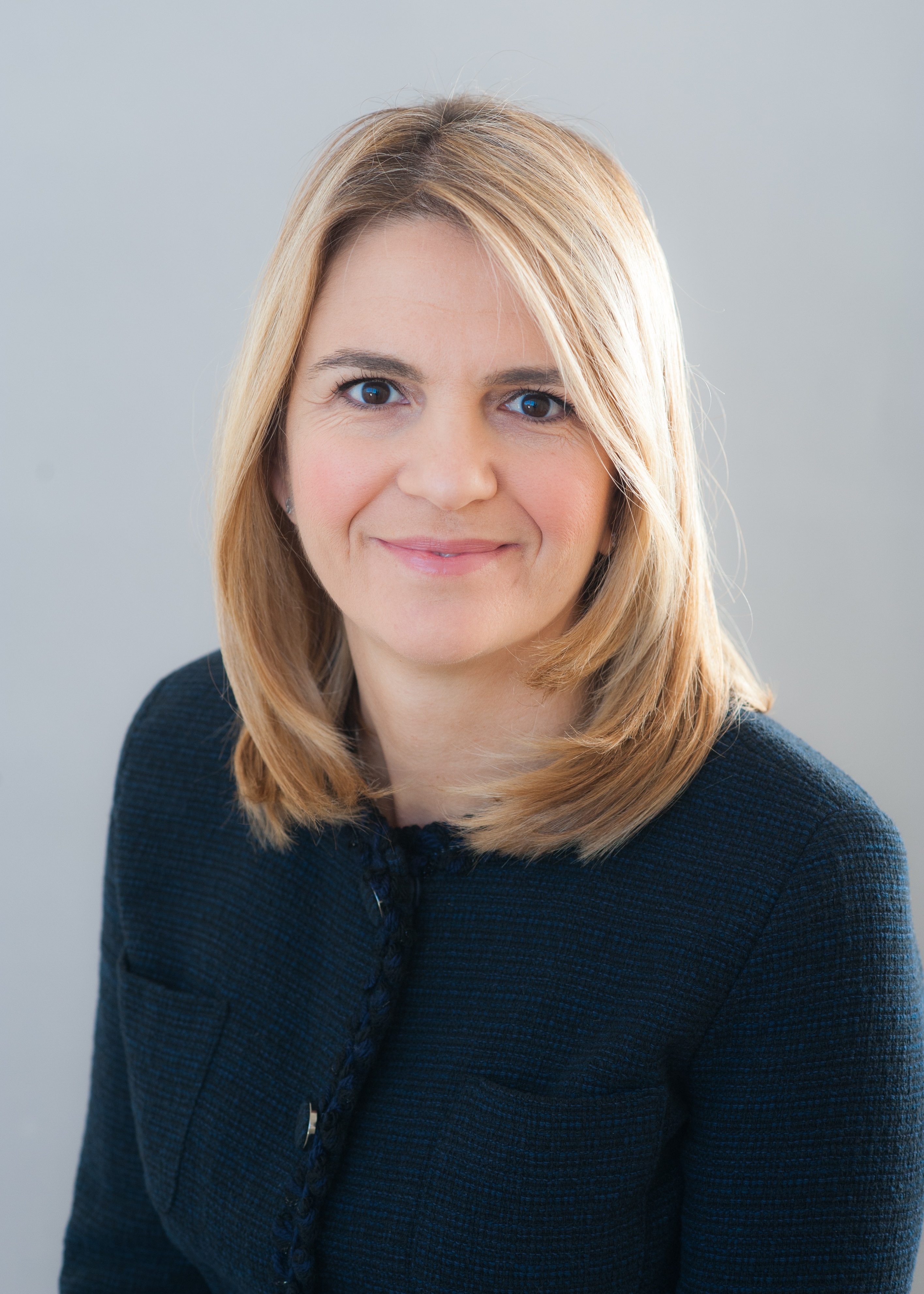}}]{Danijela Cabric}
is Professor in Electrical and Computer Engineering at University of California, Los Angeles. She earned MS degree in Electrical Engineering in 2001, UCLA and Ph.D. in Electrical Engineering in 2007, UC Berkeley, Dr. Cabric received the Samueli Fellowship in 2008, the Okawa Foundation Research Grant in 2009, Hellman Fellowship in 2012, the
National Science Foundation Faculty Early Career Development (CAREER) Award in 2012, and the Qualcomm Faculty award in 2020 and 2021. She served as an Associate Editor of IEEE Transactions of Cognitive Communications and Networking, IEEE Transactions of Wireless Communications, IEEE Transactions on Mobile Computing and IEEE Signal Processing Magazine, and IEEE ComSoc Distinguished Lecturer. Her research interests are millimeter-wave communications, distributed communications and sensing for Internet of Things, and machine learning for wireless networks co-existence and security. She is an IEEE Fellow.
\end{IEEEbiography}

\end{document}